\documentclass[11pt]{article}
\usepackage[left=2.5cm, right=2.5cm, top=2cm, bottom=2.0cm]{geometry}
\usepackage{graphicx,upgreek}
\usepackage{graphicx}
\usepackage{bm}
\usepackage{verbatim}
\usepackage{overpic}
\usepackage[usenames,dvipsnames]{xcolor}
\usepackage[colorinlistoftodos]{todonotes}
\usepackage{soul}
\usepackage{bm,color,amssymb}
\usepackage{color,nameref}
\usepackage{hyperref}

\newcommand*{\doi}[1]{\href{http://dx.doi.org/#1}{\ (doi: #1)}}
\usepackage{cite}
\usepackage{amsmath}
\usepackage{amssymb}
\usepackage{mathtools}
\usepackage{subcaption}
\usepackage{ulem}
\usepackage[numbers]{natbib}
\newcommand{\intg}{\int d \mathbf{x}}

\newcommand{\avg}[2] {\left \langle #1 \right \rangle_{\text{#2}}}

\begin{document}  
\bibliographystyle{unsrtnat}
\begin{center}
\begin{Large}
{\bf  Active Inter-cellular Forces in Collective Cell Motility}\\
~\\
\end{Large}

Guanming Zhang$^1$, Romain Mueller$^1$, Amin Doostmohammadi$^2$ and Julia M. Yeomans$^1$\\
~\\
$^1$The Rudolf Peierls Centre for Theoretical Physics, University of Oxford, Parks Road, Oxford OX1 3PU, UK\\
$^2$The Niels Bohr Institute, University of Copenhagen, Blegdamsvej 17, 2100 Copenhagen, DK\\
\end{center}
~\\

\begin{abstract}
The collective behaviour of confluent cell sheets is strongly influenced both by polar forces, arising through cytoskeletal propulsion, and by active inter-cellular forces, which are mediated by interactions across cell-cell junctions. We use a phase-field model to explore the interplay between these two contributions  and compare the dynamics of a cell sheet when the polarity of the cells aligns to (i) their main axis of elongation, (ii) their velocity, and (iii) when the polarity direction executes a persistent random walk.
In all three cases, we observe a sharp transition from a jammed state (where cell rearrangements are strongly suppressed) to a liquid state (where the cells can move freely relative to each other) when either the polar or the inter-cellular forces are increased.
In addition, for case (ii) only, we observe an additional dynamical state, flocking (solid or liquid), where the majority of the cells move in the same direction. The flocking state is seen for strong polar forces, but is destroyed as the strength of the inter-cellular activity is increased.
\end{abstract}

\section{Introduction}

The ways in which cells move collectively have implications for a vast range of cellular processes, from embryogenesis and morphogenesis \cite{Heisenberg2013} to wound healing \cite{brugues2014forces}, cancer metastasis \cite{friedl2012classifying} and the viability of medical implants \cite{Hu2019implants}. However, the mechanisms behind the collective motion of cells are still unclear even though considerable theoretical understanding has been recently achieved using two-dimensional simplified model systems of confluent cell layers~\cite{ladoux2017mechanobiology,alert2020physical}.
In particular, the importance of active inter-cellular forces and their interplay with cell polarity remain largely unknown.
Here we use a phase-field model of a cell sheet to investigate these questions and show how different forcing can lead to a range of collective cell behaviours: jamming, a liquid state reminiscent of active turbulence, and flocking.
By doing so, we build towards a framework for interpreting experiments on the ways cells move.

The driving forces behind a single cell moving across a flat surface are well understood. The cell is controlled by directional actin filaments, which can continuously polymerise and depolymerise to form lamellopodia, protrusions that push the cell forwards \cite{mitchison1996actin,alberts2002molecular}. To advance, the cell needs to push against the substrate, and to do this effectively it creates focal adhesions, which are mechanical links between internal actin bundles and the external surface \cite{sarangi2016coordination}. As it moves, the cell tends to polarise and elongate in the direction of motion.
This means that a minimal physical picture of single cell motility can be obtained by considering only a net force in the direction of the cell polarity.

When cells form a confluent layer the inter-cellular contacts mediate a striking change in the dynamics \cite{farhadifar2007influence,friedl2009collective}.  As long as the cells are not too tightly packed, they continually move around within the cluster, with flow patterns that show high vorticity and localised bursts of high velocities \cite{petitjean2010velocity}, reminiscent of the  active turbulence that characterises dense active nematic materials \cite{Doostmohammadi2018active}. Lamellopodium formation is usually suppressed in confluent cell layers and, although cryptic lamellopodia  can sometimes appear beneath the cells, these appear to have little effect on the cell dynamics.  However, in the presence of an external stimulus, e.g.~a wound or a variation in substrate stiffness, the cell layer can flock, with the cells all moving together in a preferred direction \cite{farooqui2005multiple}. For example, adding RAB5A to a confluent cell layer leads to flocking, with the cells moving coherently in a preferred direction, and about five times faster than in the unperturbed state \cite{palamidessi2019unjamming}. Another example is the epithelial layer of follicle cells
lining the {\it Drosophila} egg chamber that rotate coherently around the chamber as it lengthens \cite{Cetera2014}. These results show that inter-cellular interactions are key in controlling how cells move collectively. In particular, active interactions mediated by E-cadherin junctions between the cells have been shown to be important in collective cell motion during wound healing~\cite{brugues2014forces}, and in collective cell guidance ~\cite{tambe2011collective}.

In silico models of cell motility have an important role to play in unravelling the relative contributions of the various forces that can determine how cells move collectively \cite{bi2015density,alert2020physical}. Numerical approaches have included cellular Potts  \cite{graner1992simulation}, vertex \cite{farhadifar2007influence}, continuum, and phase-field models \cite{lober2015collisions}. 
The crossover from jammed dynamics to flocking has been reproduced in a vertex model by including an alignment between the polarisation of a cell and its velocity. 
Similarly, agent-based models with polarisation-velocity alignments and repulsive interactions lead to a flocking motion \cite{giavazzi2017giant,szabo2006phase}. The role and impact of active intercellular forces has, however, so far remained unexplored.

Here we use a phase-field approach which allows single cell properties, such as cell deformability or cell polarity, and inter-cellular interactions to be varied independently. We focus on two active forces: the polar force on each cell and the active inter-cellular forces. Varying them independently allows us to compare the conditions that lead to three different states: jamming, liquid and flocking.
Hence, our model unifies a wide range of collective cell dynamics in a single coherent description. In the next section we describe the details of the numerical model.  In section~3  we present our results, considering in turn the unjamming transition, cell shapes and the liquid phase, and then cell velocities and flocking. Section~4 discusses our conclusions and suggests ideas for future simulations and experiments.

\section{The phase field model}

We model a monolayer of cells using a coarse-grained, phase field approach that resolves individual cells and the forces between them, but not the internal machinery of the cell. The phase field method for a single cell is reviewed in \cite{lober2014modeling,ziebert2011model,ziebert2013effects}; it has been used to study the contact inhibition of locomotion and collisions in binary cell systems \cite{camley2014polarity,lober2015collisions}, the effect of the stiffness mismatch between single cancer cells and normal cells in metastasis \cite{palmieri2015multiple} and the role of cell overlaps in the solid-liquid transition of a cell sheet \cite{loewe2019solid}.

In this setting, the extent of each cell $i$ is defined by a phase field $\varphi_i$.  Each phase field $\varphi_i(\mathbf{x})$ moves with velocity $\mathbf{v}_i(\mathbf{x})$ according to the equation of motion
\begin{equation}
\label{eq:dynamics}
\partial_t \varphi_i(\mathbf{x}) + \mathbf{v}_i (\mathbf{x}) \cdot \nabla \varphi_i(\mathbf{x} ) = -\frac{\delta {\mathcal{F}}}{\delta \varphi_i (\mathbf{x})},
\end{equation}
where ${\mathcal F}$
is the total free energy of the cell layer.
As cell Reynolds numbers are typically $\sim 10^{-4}$, we assume overdamped dynamics and relate the velocity of each point of a cell in Eq.~(\ref{eq:dynamics}) to the force acting on it by
\begin{equation}
\label{eq:force_balance}
\xi \mathbf{v}_i (\mathbf{x}) =\mathbf{f}^{\text{tot}}_i (\mathbf{x}),
\end{equation}
where $\xi$ is a friction coefficient and $\mathbf{f}^{\text{tot}}_i$ is the total force exerted on point $i$.
We consider three contributions to the force acting on each cell, $\mathbf{f}_i^{\text{tot}} = \mathbf{f}_i^{\text{passive}} + \mathbf{f}_i^{\text{pol}} + \mathbf{f}_i^{\text{int}} $, a passive force accounting for equilibrium cell properties such as compressibility, deformability, and the strength of cell-cell repulsion, an active polar force acting along the direction of the cell polarization, and a force describing inter-cellular stresses that actively remodel cell-cell contacts.

\subsection{Passive force}

Passive forces stem from an effective free energy; if a cell deviates from equilibrium, the passive thermodynamic force will drive it towards a lower free energy state.
This can be expressed as \cite{cates2018Theories}
       $$\mathbf{f}_i^{\text{passive}}(\mathbf{x}) = \frac{\delta \mathcal{F}}{\delta \varphi_i} \nabla \varphi_i,$$
where we write the free energy as  $\mathcal{F} = \mathcal{F}_{\text{CH}} + \mathcal{F}_{\text{area}} + \mathcal{F}_{\text{rep}} + \mathcal{F}_{\text{adh}}$ \cite{mueller2019emergence}.

The first contribution, defined as
\begin{equation}
{\mathcal F}_{\text{CH}}=\sum_i \frac{\gamma}{\lambda} \int d \mathbf{x} \left\{ 4 \varphi_i^2(1-\varphi_i)^2+\lambda^2(\nabla \varphi_i)^2\right\},
\end {equation}
is a Cahn-Hilliard free energy that  encourages $\varphi_i$ to take values $1$, which we choose to correspond to the inside of the cell $i$, or $0$, which denotes the region outside the cell. The cell boundary is located at $\varphi = 1/2$ and has width ${\mathcal{O}}(\lambda)$, set by the gradient term, and ${\gamma}/{\lambda}$ is an energy scale. The Cahn-Hilliard free energy controls the deformability of the individual cells through the parameter $\lambda$. 

In addition to the cell deformability, we account for the cell compressibility using a free energy term
\begin{equation}
{\mathcal F}_{\text{area}}=\sum_i \mu \left\{ 1-\frac{1}{\pi R^2} \int d \mathbf{x} \; \varphi_i^2 \right \}^2,
\end {equation}
which imposes a soft constraint, of strength $\mu$, restricting the area of each cell to $\pi R^2$. Therefore, in the absence of any active forces or cell-cell interactions, the cells will relax to circles. Note that since this free energy contribution is summed over all cells, the model monolayer is compressible.

The passive contributions to cell-cell interactions are introduced through repulsion and adhesion free energies as
\begin{equation}
{\mathcal F}_{\text{rep}}=\sum_i \sum_{j \neq i} \frac{\kappa}{\lambda} \int d \mathbf{x}\; \varphi_i^2 \varphi_j^2,
\end {equation}
which penalises overlap between cells with an energy scale ${\kappa}/{\lambda}$ and 
\begin{equation}
{\mathcal F}_{\text{adh}}=\sum_i \sum_{j \neq i} \omega\lambda \int d \mathbf{x}\; \nabla \varphi_i \cdot \nabla \varphi_j,
\label{adhesion}
\end {equation}
which is a term favouring cell-cell adhesion. Note that $ \lambda \intg{\nabla \varphi_i \cdot \nabla \varphi_j }$ measures the  length of contact line between two cells and $\omega / \lambda$ is an energy scale.

If no active forces or external forces are acting, the cell layer will relax to a minimum of the total free energy. The global minimum is for cells in a confluent layer to be static, identical hexagons. 
 However, cells are active, self-motile systems, driven out of equilibrium by chemical processes. Therefore we next describe how we account for active inter-cellular forces and active polar forces in the phase field model.

\subsection{Active inter-cellular forces\label{sec:actnem}}

Force transduction at cell-cell junctions, is mediated by E-cadherin bonds which actively modify the actomyosin networks inside the connecting cells leading to cell deformations~\cite{Ng2014mapping}. To model the active inter-cellular forces we calculate the deformation tensor of each cell,
\begin{equation}
\label{eq:deviation_tensor}
{\mathcal D}_i = - \int d\mathbf{x} \left\{\nabla \varphi_i \nabla \varphi_i^T - \frac{1}{2} \mbox{Tr}( \nabla \varphi_i \nabla \varphi_i^T )\right\},
\end{equation}
and define an associated stress field due to all the cells,
\begin{equation}
\sigma_{D} = -\zeta \sum_i \varphi_i(\mathbf{x}) {\mathcal D}_i,
\end{equation}
where $\zeta$ measures the strength of the forcing.
The active inter-cellular force density is then
\begin{equation}
\mathbf{f}(\mathbf{x})^{\text{int}} = \nabla \cdot \sigma_{D}.
\end{equation}
This is constant within the cells but acts at junctions. 
We take $\zeta>0$ where, 
for an individual cell, the force density further elongates the cell along its long axis (Fig.~\ref{fig:phasediagrama}).   
For a collection of cells, the active inter-cellular force also enhances any parallel alignment of neighbouring cells.

The deformation tensor can be written in terms of its eigenvalues and eigenvectors \cite{mueller2019emergence} as
\begin{equation}
\label{eq:nematic_tensor}
{\mathcal D}_i = S_i(\mathbf{n}_i \mathbf{n}_i^T - {\mathcal I}/2),
\end{equation}
where $\mathbf{n}_i$, the direction of the elongation axis of the cell, is the normalised eigenvector of ${\mathcal D}_i$ which corresponds to its largest eigenvalue and $S_i$ is half the difference between its eigenvalues. A measure of the deformation of a cell is then 
\begin{equation}
    \label{eq:deformation}
    D_i = \frac{1}{2}S_i = \sqrt{\mathcal{D}_{xx,i}^2 + \mathcal{D}_{xy,i}^2},
\end{equation} 
where $\mathcal{D}_{xx,i}$ and $\mathcal{D}_{xy,i}$ are the $xx$ and $xy$ components of the deformation tensor.

\subsection{Active polar force\label{sec:polalign}}

\begin{figure}[t!] 
 	\centering 
		\begin{subfigure}[b]{0.15\textwidth}
		\includegraphics[width=70pt]{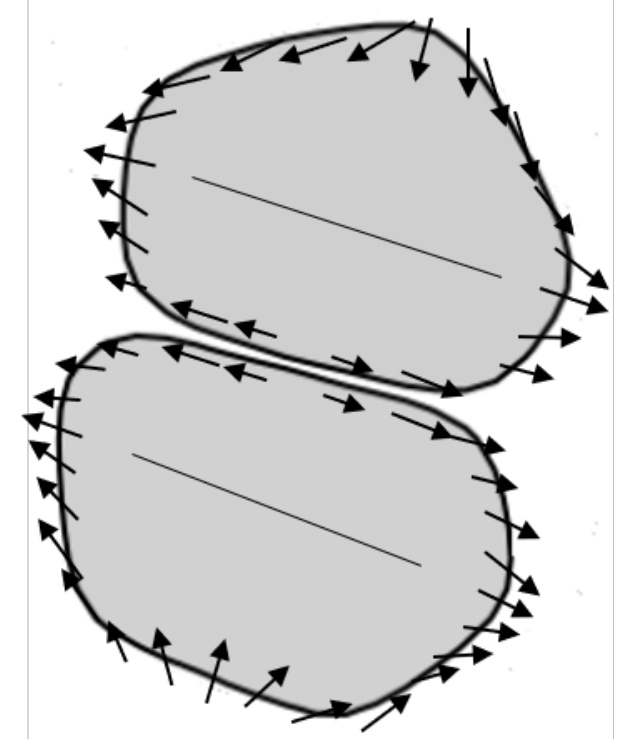}
	\caption{}
		\label{fig:phasediagrama}
		\centering 
 	\includegraphics[width=70pt]{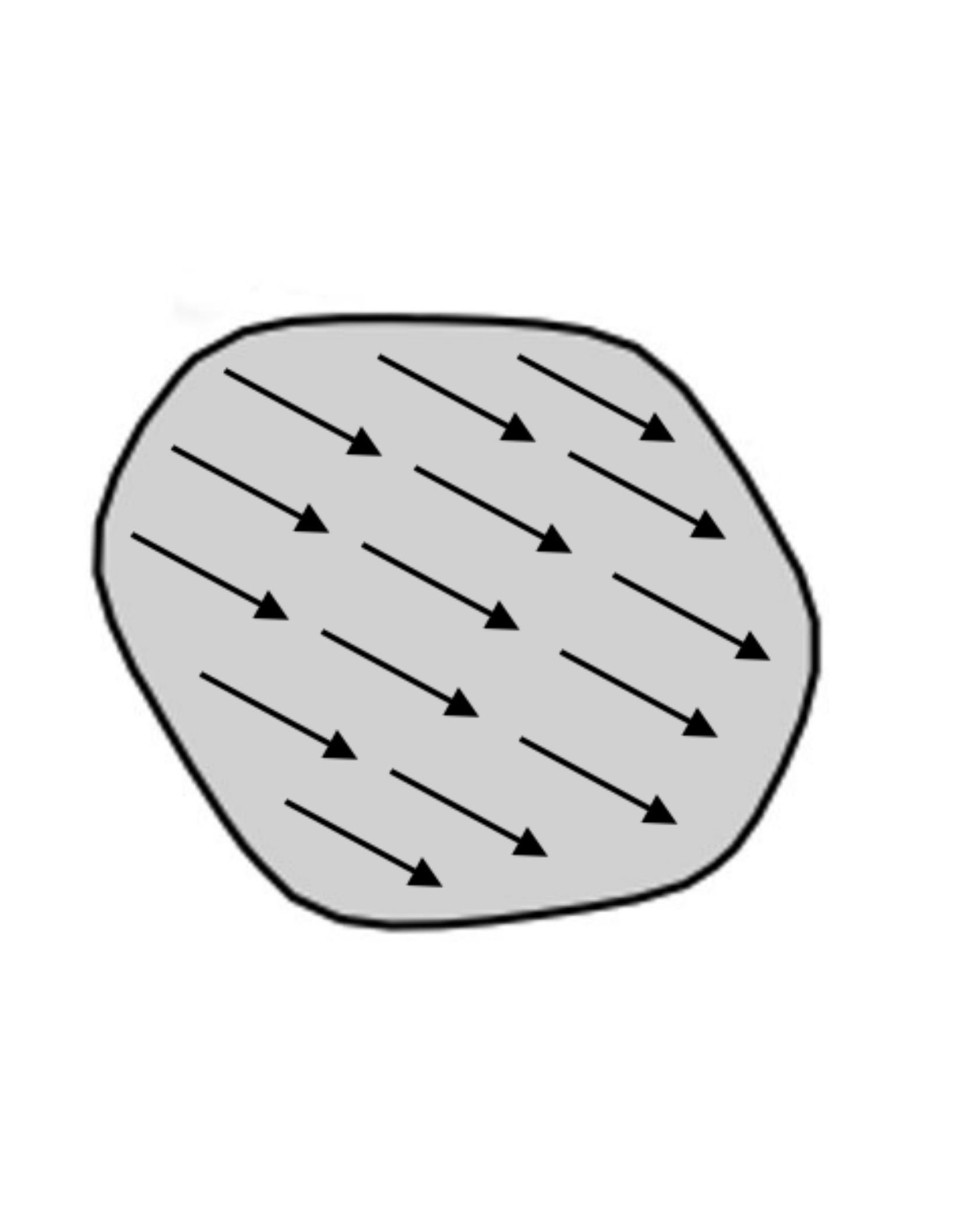}			
		\caption{}
			\label{fig:phasediagramb}
	\end{subfigure}
 \begin{subfigure}[b]{0.4\textwidth}
 	\centering 
	\includegraphics[width=200pt]{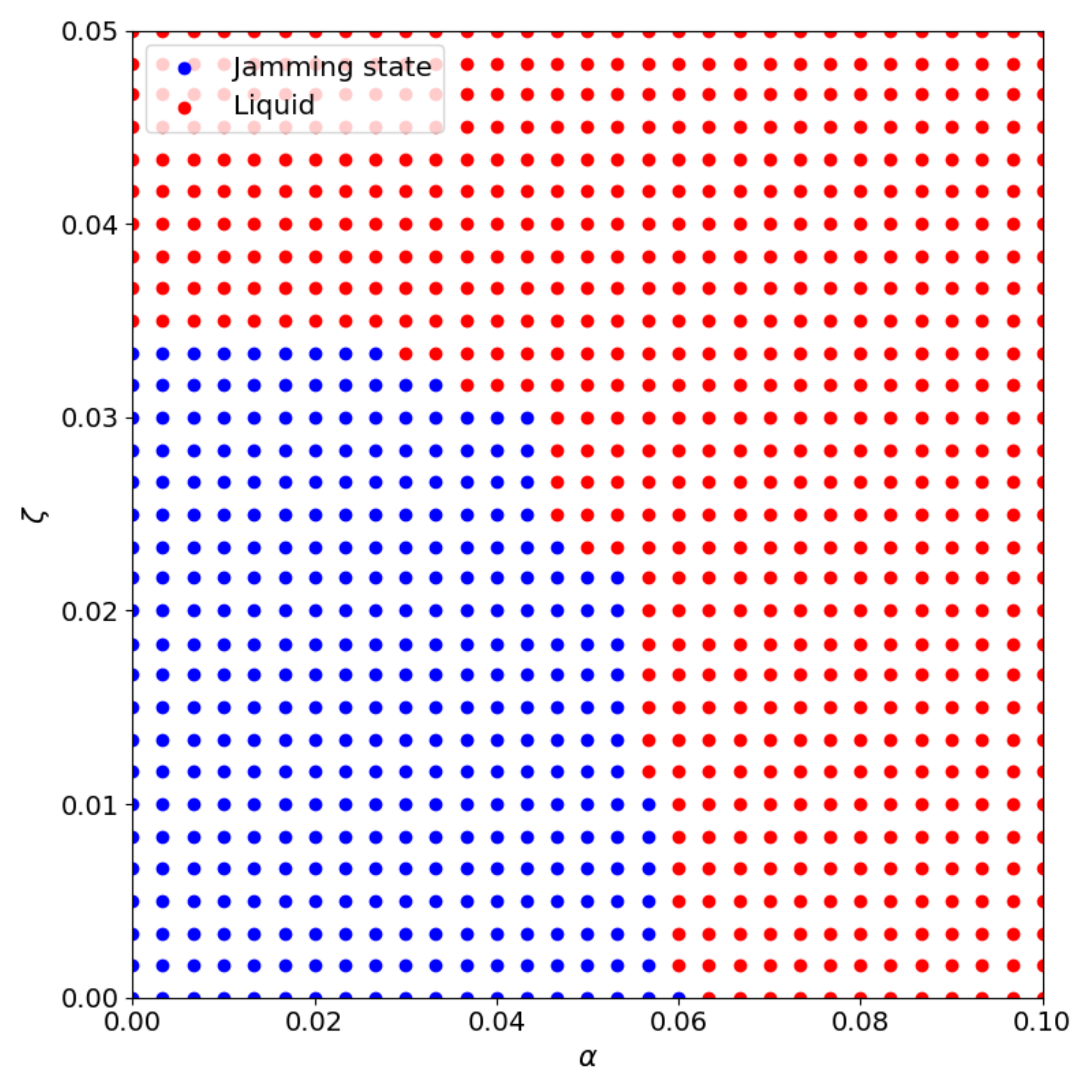}
	\caption{}
		\label{fig:phasediagramc}
\end{subfigure}
\begin{subfigure}[b]{0.4\textwidth}
 	\includegraphics[width=200pt]{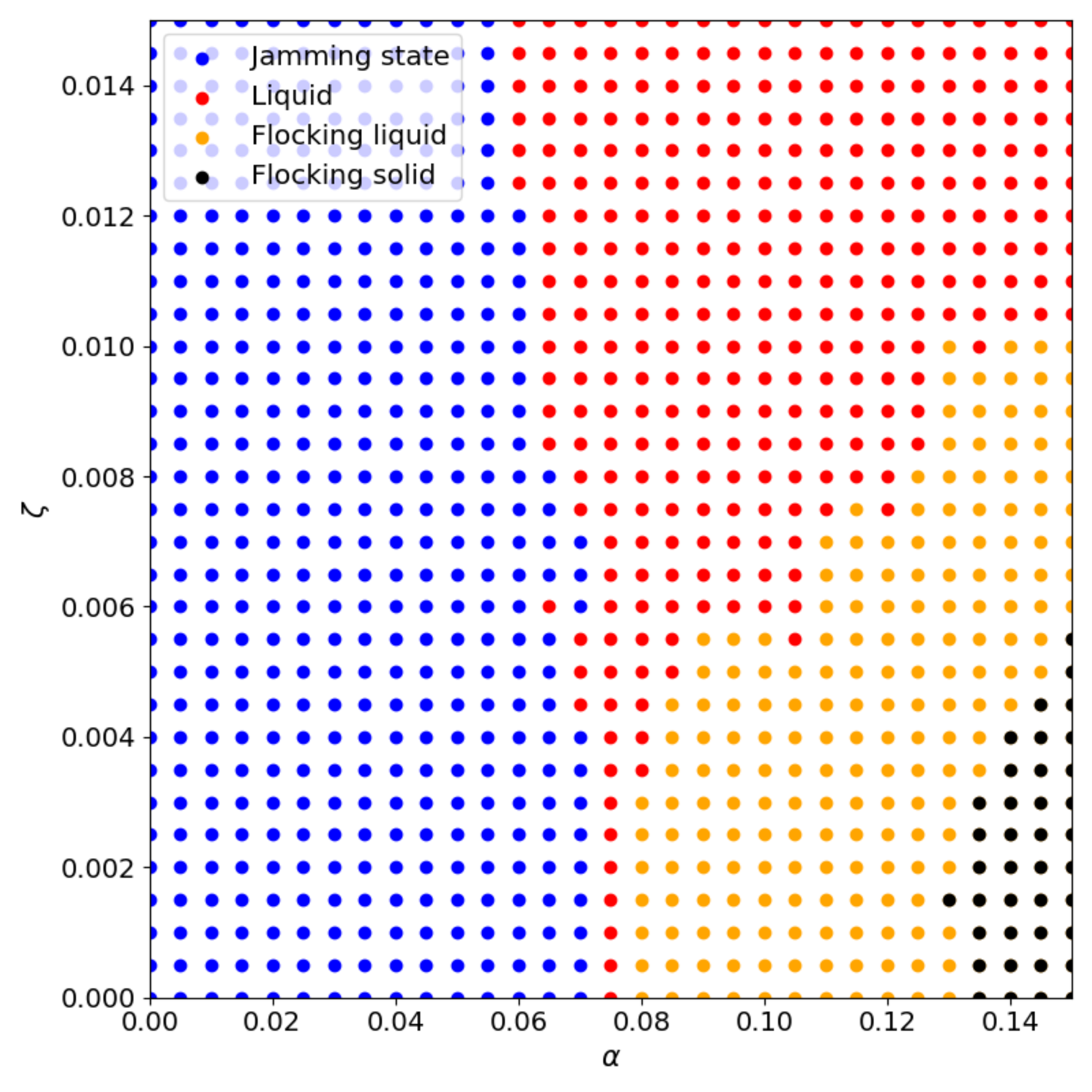}
	\caption{}
		\label{fig:phasediagramd}
	\end{subfigure}
 	\caption{{\bf Stability diagrams:} Modelling the active force densities acting on a cell: (a)  inter-cellular forces, (b) polar forces due to actomyosin driving. In (a) the solid lines denote the long axis of the cell. Stability diagram for (c) case {\it shape} (polarisation aligns to long axis of the cell; case {\it Brownian} is very similar) (d) case {\it velocity} (polarisation aligns to the velocity of the cell) for $J_{\text{pol}} = 0.1$ and $\omega = 0.005$. Jammed states are identified by a rearrangement rate $<$ 0.15. Liquid states have a rearrangement rate  $>$ 0.15. Flocking states have a Viscek order parameter $V_a > 0.15$. Note the different scales in (c) and (d).}	
 \end{figure}

The active polar contribution models the force acting on an individual cell due to actin treadmilling. We associate each cell with a polarisation vector $\mathbf{p}_i = (\cos{\theta_i},\sin{\theta_i})$, and define the polar force to act in the direction of the cell polarisation and to be uniformly distributed over the cell  (Fig.~\ref{fig:phasediagramb}), 
\begin{equation}
 \mathbf{f}_i^{\text{pol}}(\mathbf{x}) = \alpha \varphi_i(\mathbf{x})\mathbf{p}_i.
 \end{equation}
The strength of the polar force generated by individual cells is controlled by the coefficient $\alpha$.
The most physical way to choose the direction of the polarisation vector $\mathbf{p}_i$ is unclear~\cite{alert2020physical}. Therefore we compare the following possibilities with the aim of uncovering any differences in behaviour that might be observable in experiments:
\begin{itemize}
	\item{\bf aligning the polarisation with the velocity of the cell ({\it velocity})}\\
	~\\
	We assume that  $\mathbf{p}_i$ relaxes towards the centre of mass velocity of cell $i$, $\mathbf{v}_i^{\text{COM}} = \int d\mathbf{x} \ \mathbf{v}_i(x)/A_i$, where the area of cell $i$ is defined as $A_i = \int d\mathbf{x} \ \varphi_i(x)^2$. The alignment dynamics reads
	\begin{equation}
	\label{eq:pol_fix_velocity}
	\partial_t \theta_i = -J_{\text{pol}} |A_i \mathbf{v}_i^{\text{COM}}| \angle ( \mathbf{p}_i, \mathbf{v}_i^{\text{COM}}) + D_{\text{pol}}\eta
	\end{equation}
	where $\eta$ is Gaussian noise with unit standard deviation and strength $D_{\text{pol}}$, $\angle ( \mathbf{p}_i,\mathbf{v}_i^{\text{COM}})$ is the signed angle from $\mathbf{v}_i^{\text{COM}}$ to $\mathbf{p}_i$, chosen to lie in the interval $(-\pi,\pi)$ and $J_{\text{pol}}$ controls the rate of alignment. This form of the polarity alignment is motivated by considering the cell-substrate interaction which, in the over-damped dynamics of the cell motion, can favour the alignment of polarity to the cell velocity~\cite{kabla2012collective,szabo2010collective}. While some experiments have reported mis-alignment of polarity and velocity~\cite{kim2013propulsion,notbohm2016cellular}, a recent study has shown that such an alignment occurs, albeit with a well-defined time delay~\cite{peyret2019sustained}.\\
	
	\item{\bf aligning the polarisation with the cell elongation (\it{shape})}\\ 
	~\\
	We take the polarisation to relax towards $\mathbf{n}_i$, i.e.~towards the long axis of the cell, 
	\begin{equation}
	\partial_t \theta_i = -J_{\text{pol}} D_i \angle ( \mathbf{p}_i, \mathbf{n}_i) + D_{\text{pol}}\eta,
	\label{eqn:case2}
	\end{equation}
	where $\angle ( \mathbf{p}_i,\mathbf{n}_i)$ is the signed angle from $\mathbf{n}_i$ to $\mathbf{p}_i$.  
	$D_i$ measures the deformation of a cell (Eq. \ref{eq:deformation}). 
	Due to its head-tail symmetry there is a freedom for the choice of $\mathbf{n}_i$ or $-\mathbf{n}_i$. We choose the one that forms an acute angle with the polarisation($\mathbf{n}_i \cdot \mathbf{p}_i \ge 0 $).
	This form of the polarity alignment, which follows from assuming that the internal acto-myosin machinery of the cell tends to align with the direction of the cell elongation, has been employed in a number of previous models~\cite{coburn2013tactile,lober2015collisions}.
	\\

 \item{\bf active Brownian motion (\it{Brownian})}\\
 ~\\
For comparison we also simulate cell layers where the polarisation direction follows a persistent random walk,
	\begin{equation}
	\label{eq:active_brownian_motion}
	\partial_t \theta_i =  D_{\text{po}l} \ \eta .
	\end{equation}
$\eta$ is Gaussian noise with unit standard deviation and strength $D_{\text{pol}}$. This form of the polarity alignment has been used in previous models~\cite{palmieri2015multiple,henkes2019universal}. However, while the motion of cells in a tissue is undoubtedly subject to a considerable amount of noise, a purely stochastic alignment mechanism is debatable in the light of experimental studies that have directly evaluated cell polarity alignments in epithelial tissues~\cite{kim2013propulsion,notbohm2016cellular,peyret2019sustained}.

\end{itemize}

\subsection{Simulation details}
	 
We simulate the dynamics of $233$ cells of radius $R = 4$  in a periodic domain of size $108\times108$ in simulation units. Initially, cells with radius $R/2$ are placed randomly, but with the constraint that the distances of neighbouring centres $> R$. They are then relaxed for $2000$ time steps under passive dynamics to reach  confluence. The simulations then start with randomly oriented velocities of magnitude $\alpha$. 
They are run for $30000$ time steps and data is collected after $10000$ time steps. Parameter values are  $\lambda=1.5$, $\gamma=0.055$, $\kappa=0.4$, $\mu=6$, $\xi=1$ and $D_{\text{pol}}=0.005$ unless otherwise specified. 

Typically the average diameter of MDCK cells $\sim10 \mu m$,  the velocity $\sim 20 \mu  m/h $, measured by Particle Image Velocimetry, and the pressure $\sim 100 Pa$, measured by Traction Force Microscopy~\cite{saw2017topological}. For these values the simulation units can be mapped to physical units using $\Delta x \sim 1 \mu m$ as the spatial interval, $\Delta t \sim 1s$ as the time step, and $\Delta F \sim 2.5 nN$ as the force unit. The polarity alignment rate $J_{\text{pol}}$ is chosen to give alignment on the time scale taken by a cell to translate through its diameter.

\section{Results}  

In order to understand the role of active forces in collective cell dynamics we focus on the interplay between inter-cellular active stresses, that are defined based on the cell deformations (Fig.~\ref{fig:phasediagrama}), and the active polar forces, proportional to the polarisation of each cell (Fig.~\ref{fig:phasediagramb}). The strength of the former is controlled by the parameter $\zeta$, while the parameter $\alpha$ controls the latter. Our results are summarised in the stability diagrams in Fig.~\ref{fig:phasediagramc} and Fig.~\ref{fig:phasediagramd}.  In Fig.~\ref{fig:phasediagramc} the polarisation of a cell aligns to its long axis (case {\it shape}) and there is an unjamming transition as either the polar or the inter-cellular forces are increased. Active Brownian dynamics of the polarisation (case {\it Brownian}) gives a very similar stability diagram. However, if the polarisation of a cell aligns to its velocity (case {\it velocity}) we obtain the stability diagram in Fig.~\ref{fig:phasediagramc}. This shows a similar unjamming transition but also, for large polar forces, flocking where the cells move coherently in the same direction. We now describe the results in more detail. 

\subsection{Unjamming}

 \begin{figure}[h!]
 \centering
  	\begin{subfigure}[b]{0.2\textwidth}
  		\centering 
  		\includegraphics[width=90pt]{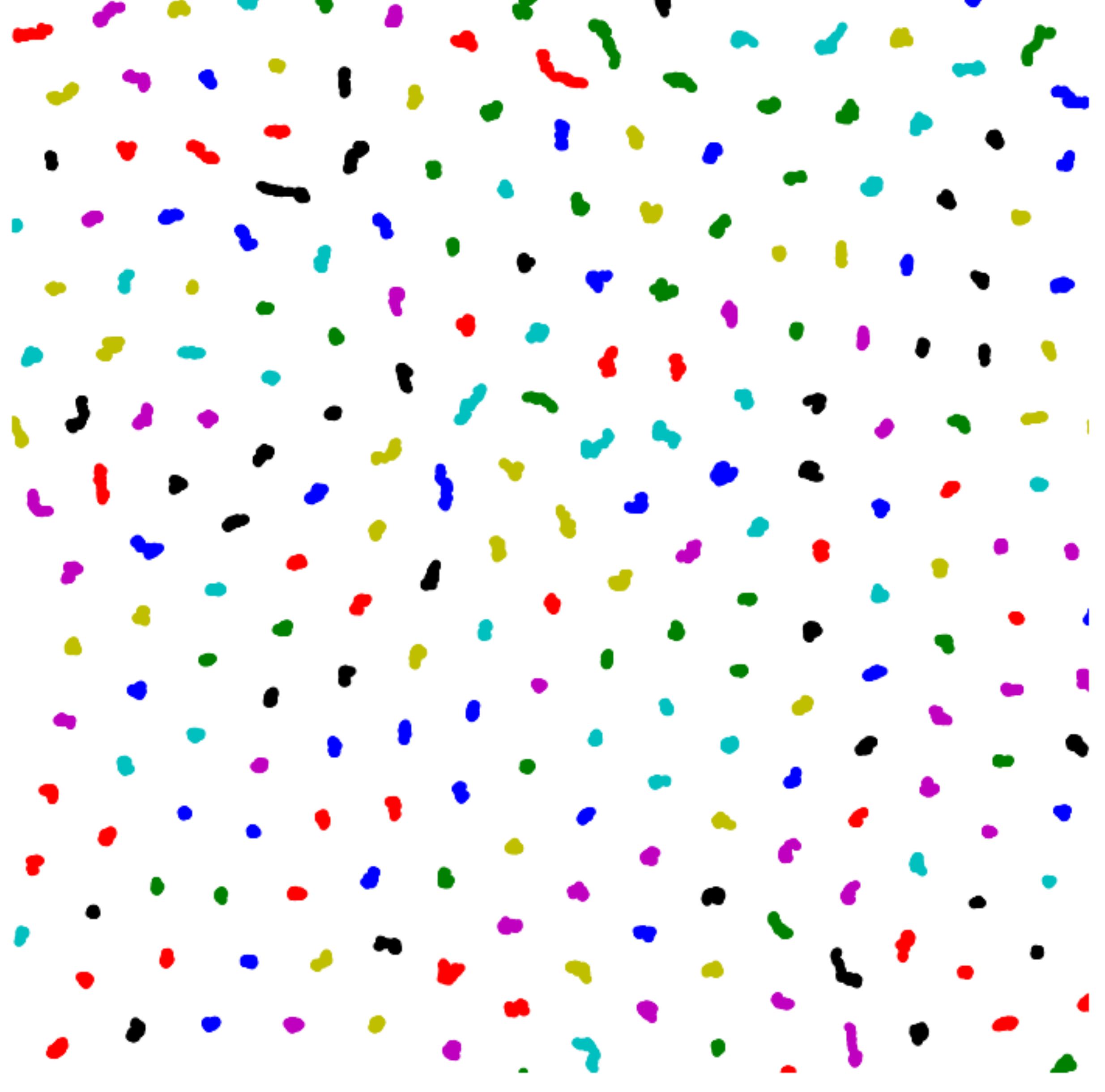}
		\caption{}
			\includegraphics[width=90pt]{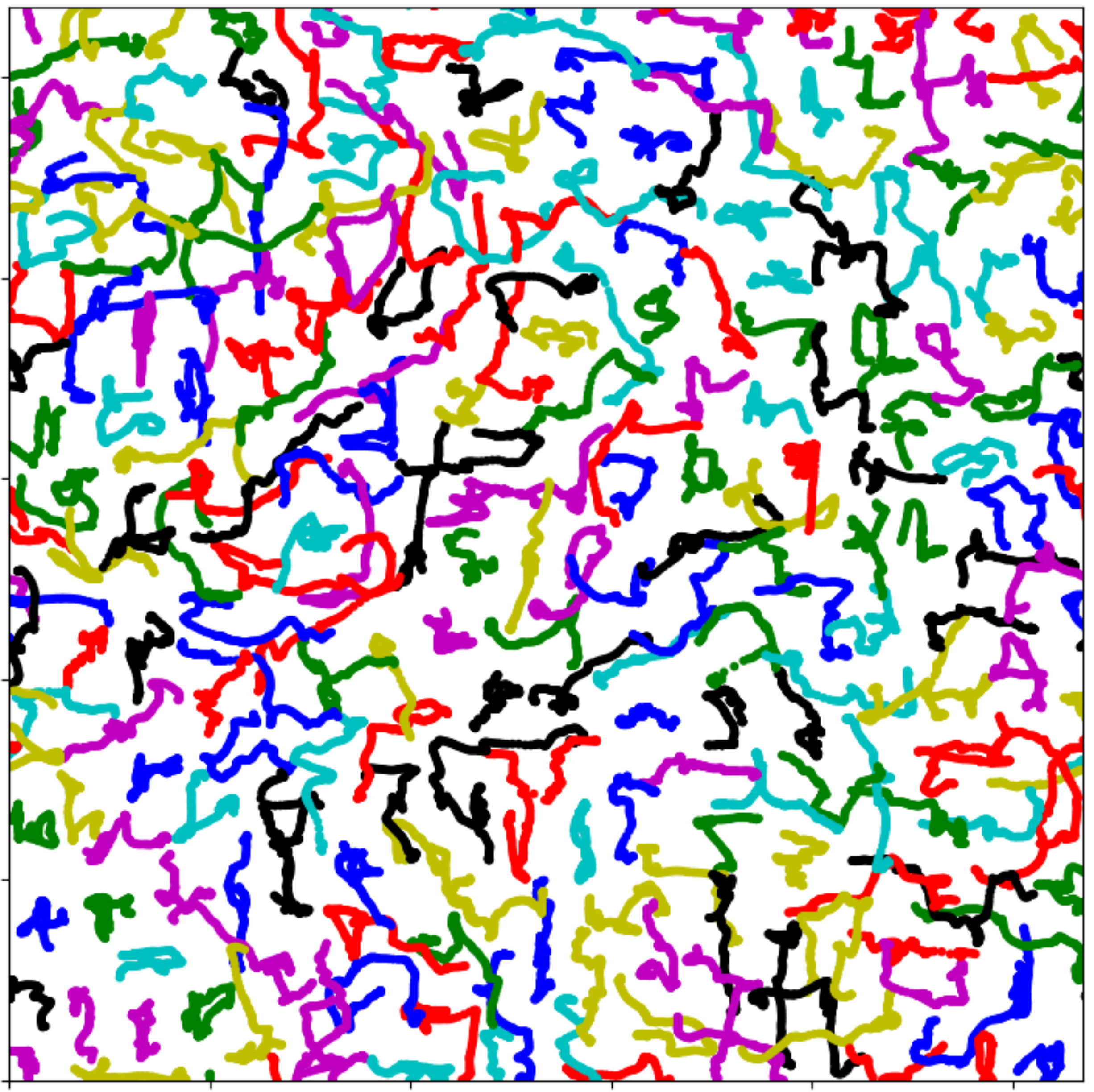}
					\caption{}
  	\end{subfigure}
  	\begin{subfigure}[b]{0.39\textwidth}	
  		\centering 
		\includegraphics[width=190pt]{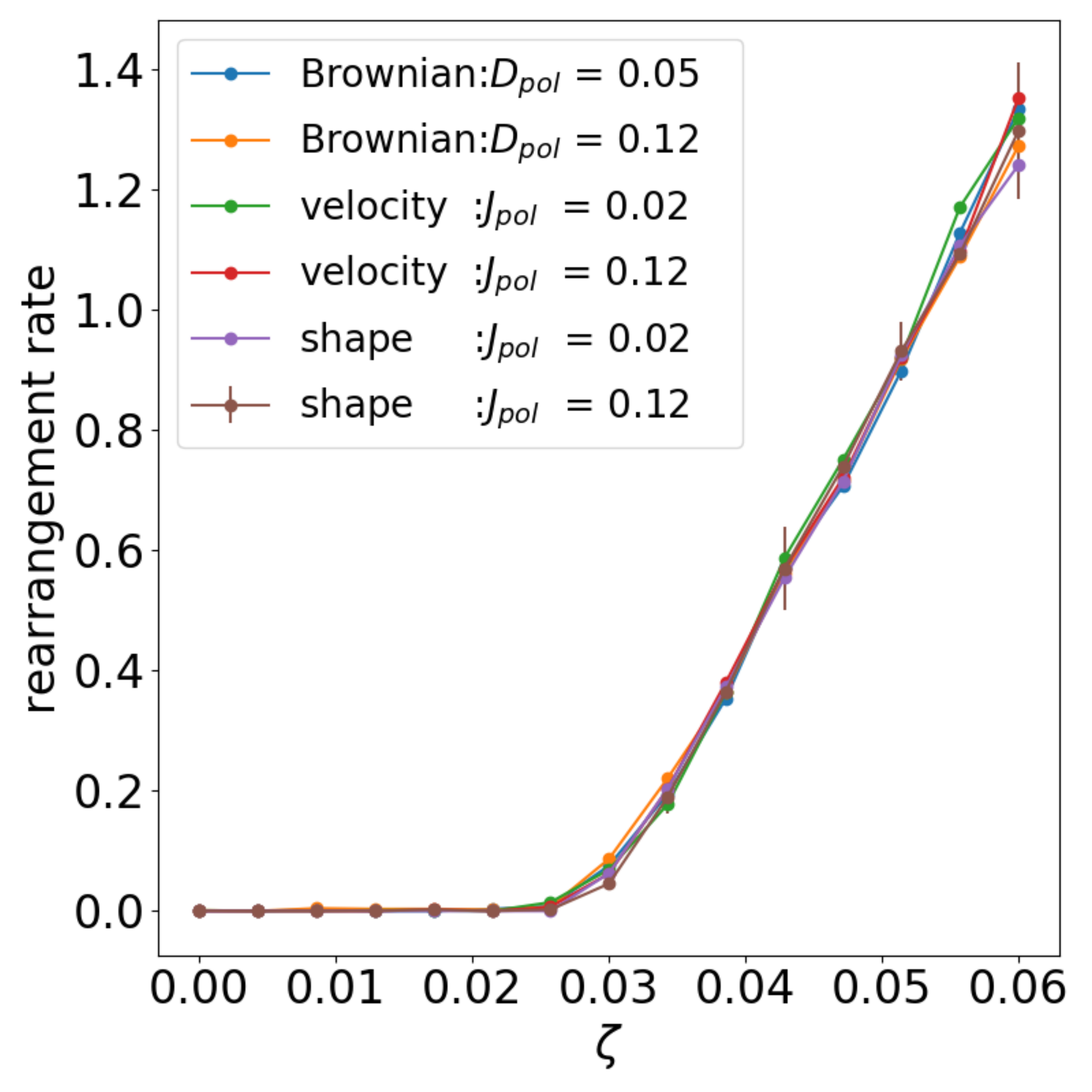}	                
		\caption{}
\end{subfigure}
				 	\begin{subfigure}[b]{0.39\textwidth}			
  		\centering 
		      \includegraphics[width=190pt]{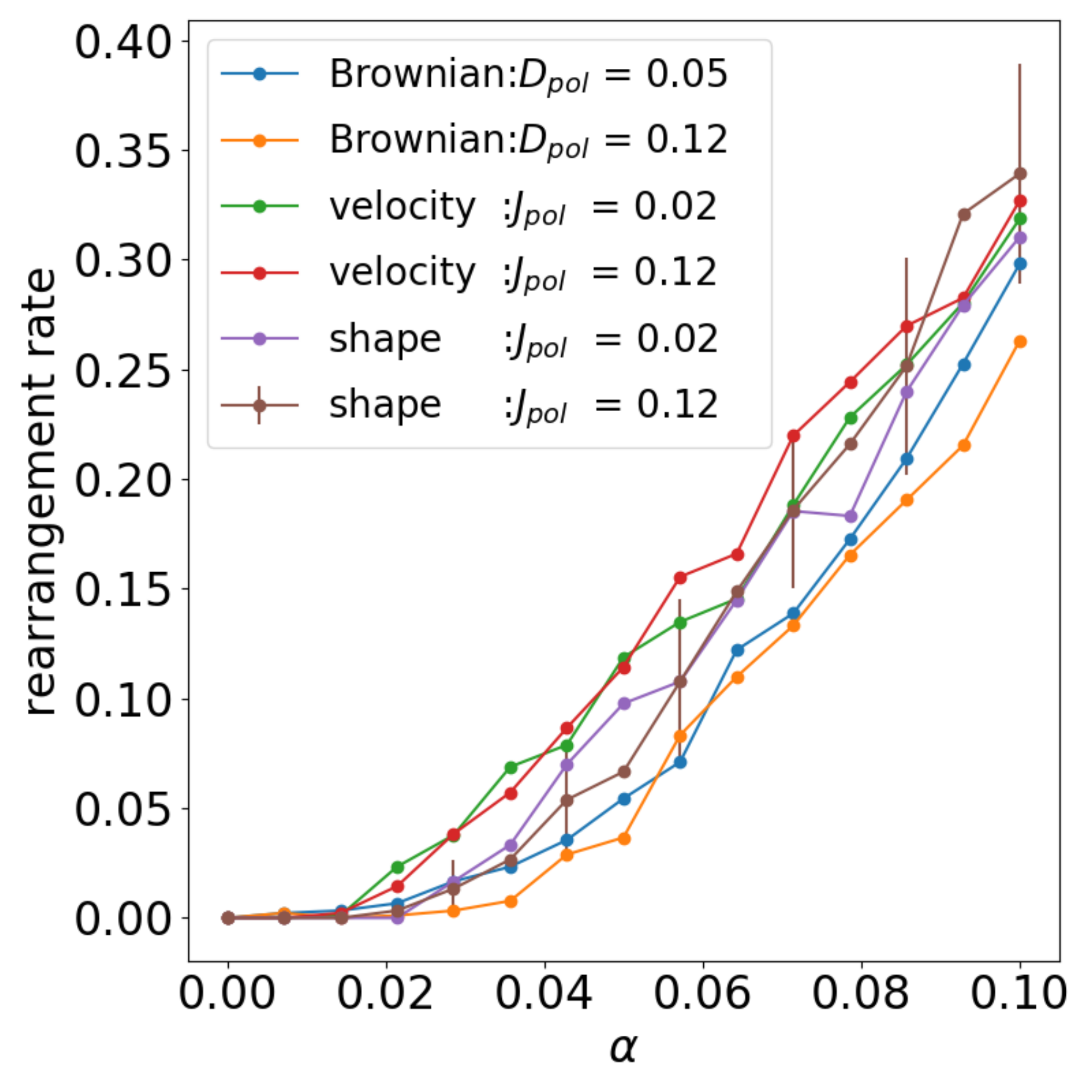}			
                          	\caption{}
				\end{subfigure}
\caption{{\bf Unjamming:} Trajectories of cells  in (a) the jammed state with $\omega = 0.005,\alpha =0.02, \zeta = 0.00$ and (b) the liquid state with $\omega = 0.005, \alpha = 0.00,\zeta = 0.04$. Trajectories are followed for $2100$ time steps and in both (a) and (b) the polarisation of a cell aligns to the long axis (case {\it shape} in Sec.~\ref{sec:polalign}).   	Rearrangement rate as a function of the strength of (c) the inter-cellular force $\zeta$ for $\alpha$=0.01, (d) the polar force $\alpha$ for $\zeta$=0.06, comparing different definitions of the polarisation alignment. $\omega=0.01$ in both frames. }
 	\label{fig:jamming-liquid}
\end{figure}

A clear feature of the cell dynamics is the crossover from a jammed state, where for small active forcing 
the cells do not move relative to each other, to a liquid state, in which rearrangements  can take place, as either the inter-cellular active stress $\zeta$ or the polar force $\alpha$ are increased. This is illustrated in Fig.~\ref{fig:jamming-liquid}(a),(b) where the trajectories of the cell centres are compared in typical jammed and liquid configurations. Similar unjamming transitions have been predicted as a function of cell shape using a vertex model \cite{bi2015density} and cell motility using a self-propelled Voronoi model \cite{bi2016motility}. 

 The transition between the two regimes can be investigated quantitatively by plotting the average number of cells that change neighbours at each time step. Fig.~\ref{fig:jamming-liquid}(c) and (d) show how this quantity changes as a function of the strengths of the  active inter-cellular  and polar forces, $\zeta$ and $\alpha$, respectively.
For a fixed value of the polar force $\alpha$, fig.~\ref{fig:jamming-liquid}(c) shows a sharp crossover between jammed and liquid dynamics as a function of $\zeta$. $\alpha$ is small so the dynamics of the polarisation alignment is irrelevant here. On the other hand, for a fixed value of the inter-cellular activity $\zeta$,
as the magnitude of the polar force is increased there is a much smoother increase in the rearrangement rate (Fig.~\ref{fig:jamming-liquid}(d)). The definition of the polarisation dynamics has only a small quantitative influence on the results: the cells unjam a little more easily if the polarisation aligns to the cell velocity. The rather sharp increase in the rearrangement rate due to increasing inter-cellular forces compared to the smoother increase due to increasing polar forces stems in the fundamentally different mechanisms by which these two sources of activity drive the monolayer out of equilibrium. For polar forces, increasing the force results in an increase in the velocity of each 
individual cell, therefore continuously enhancing cell displacement to overcome the caging effect in the jamming state. On the other hand, inter-cellular forces need to overcome the energy barrier set by the elasticity of the cells to be able to deform the cell boundaries. As such, the cell displacement only sets in once the inter-cellular forces are strong enough to deform the cells in the monolayer.

This is summarised in the stability diagram in Fig.~\ref{fig:phasediagramc}, where we choose a cell rearrangement rate of 0.15 as the boundary between the jammed and liquid phases. These results are for case {\it shape}, polar force aligned with the long axis of the cell (see Sec.~\ref{sec:polalign}). However, as is clear from the rearrangement rates plotted in Fig.~\ref{fig:jamming-liquid}, very similar behaviour is seen for all choices of the polarity alignment.

\begin{figure}[h]
 	\centering 
		\begin{subfigure}[t]{0.3\textwidth}
 	\includegraphics[width=130pt]{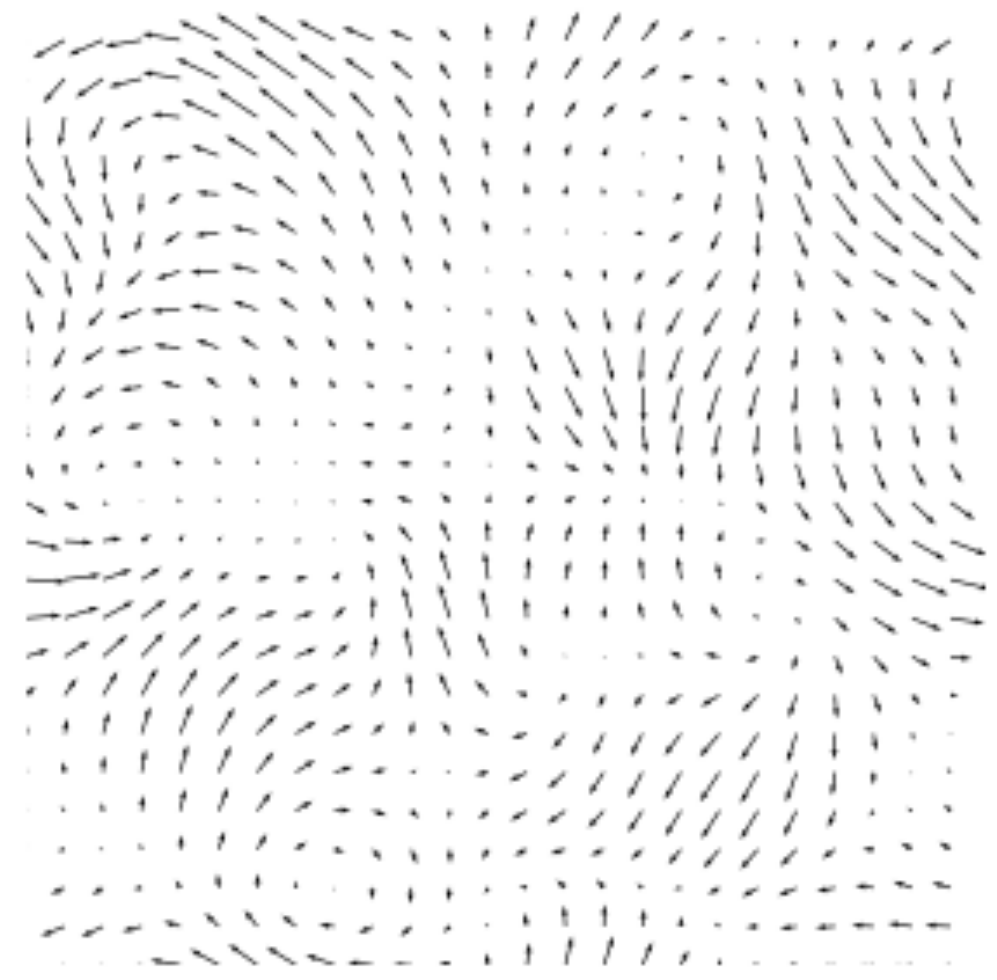}
 	\caption{}
		\label{fig:velocity-distribution-polar}
		    \includegraphics[width=130pt]{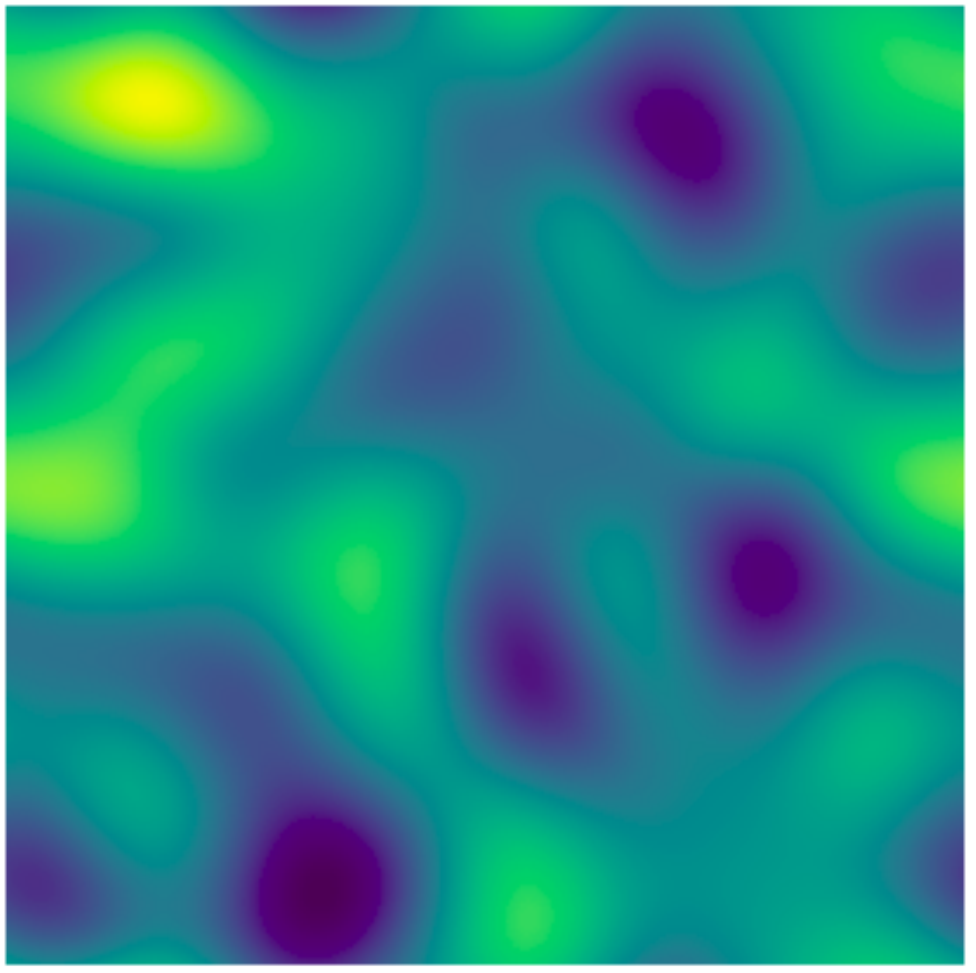}
		\caption{}
\end{subfigure}
	\begin{subfigure}[t]{0.33\textwidth}
		\centering 	
		\includegraphics[width=155pt]{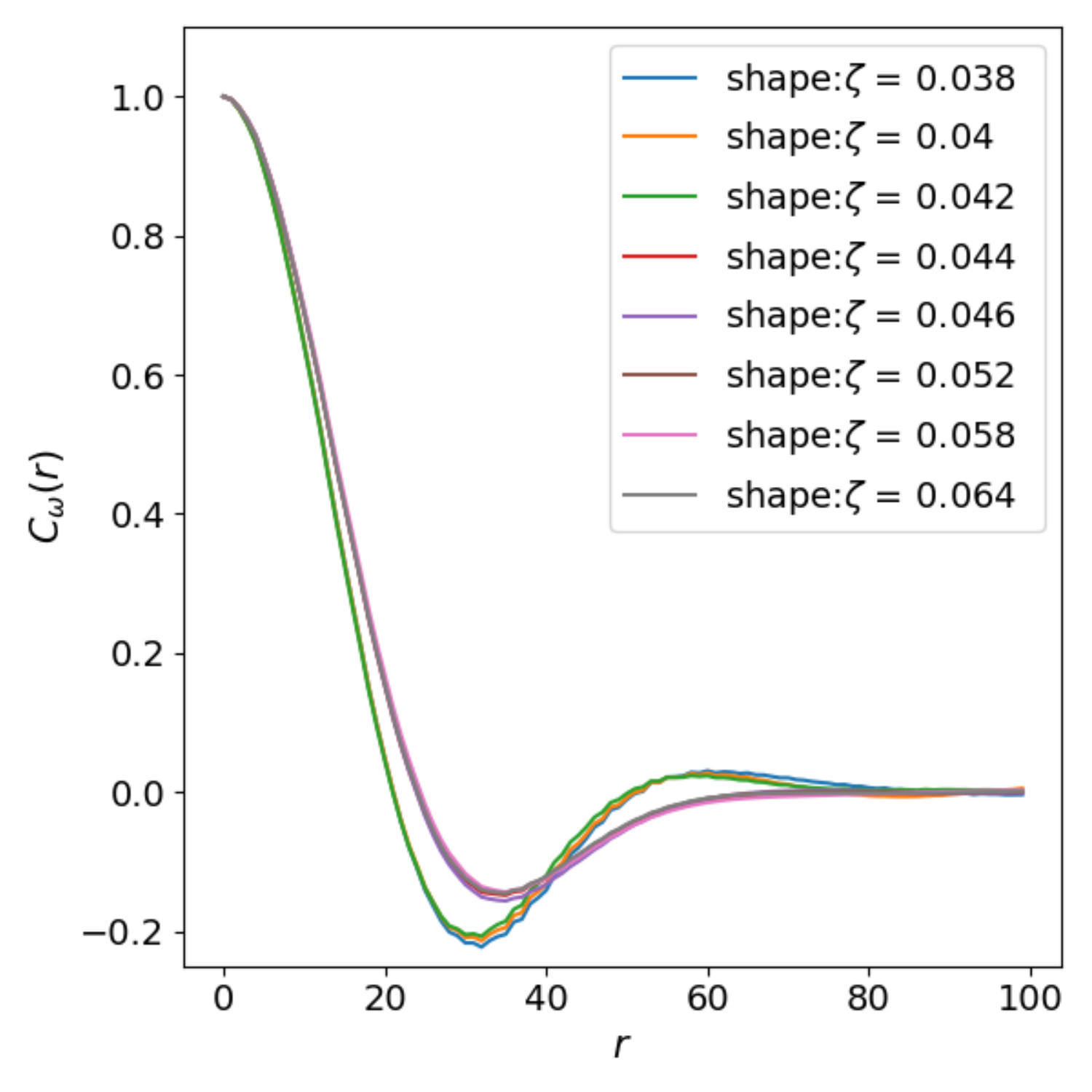}
		\caption{}
		\includegraphics[width=155pt]{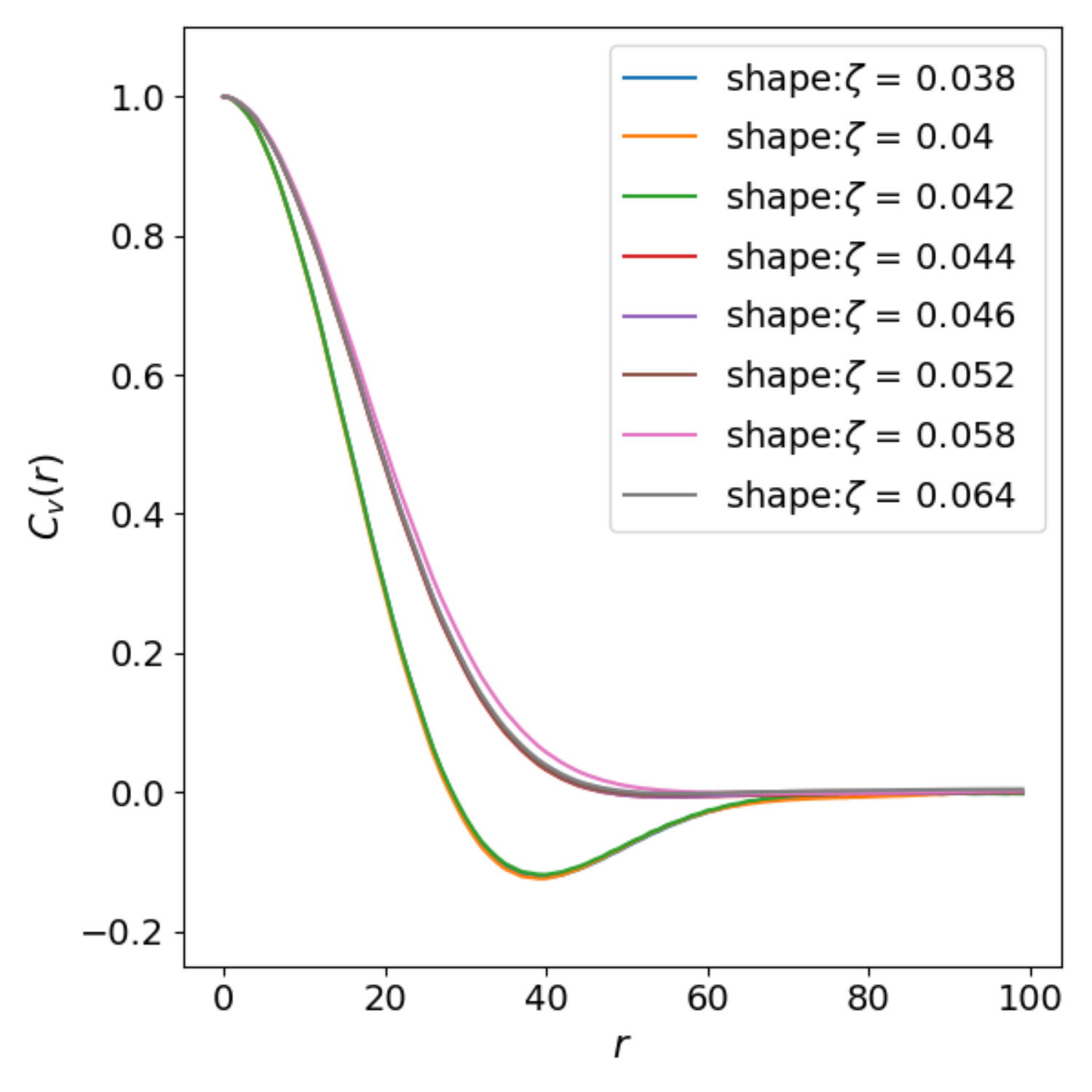}
		\caption{}
			\end{subfigure}	
			\begin{subfigure}[t]{0.33\textwidth}
		\centering 
		\includegraphics[width=155pt]{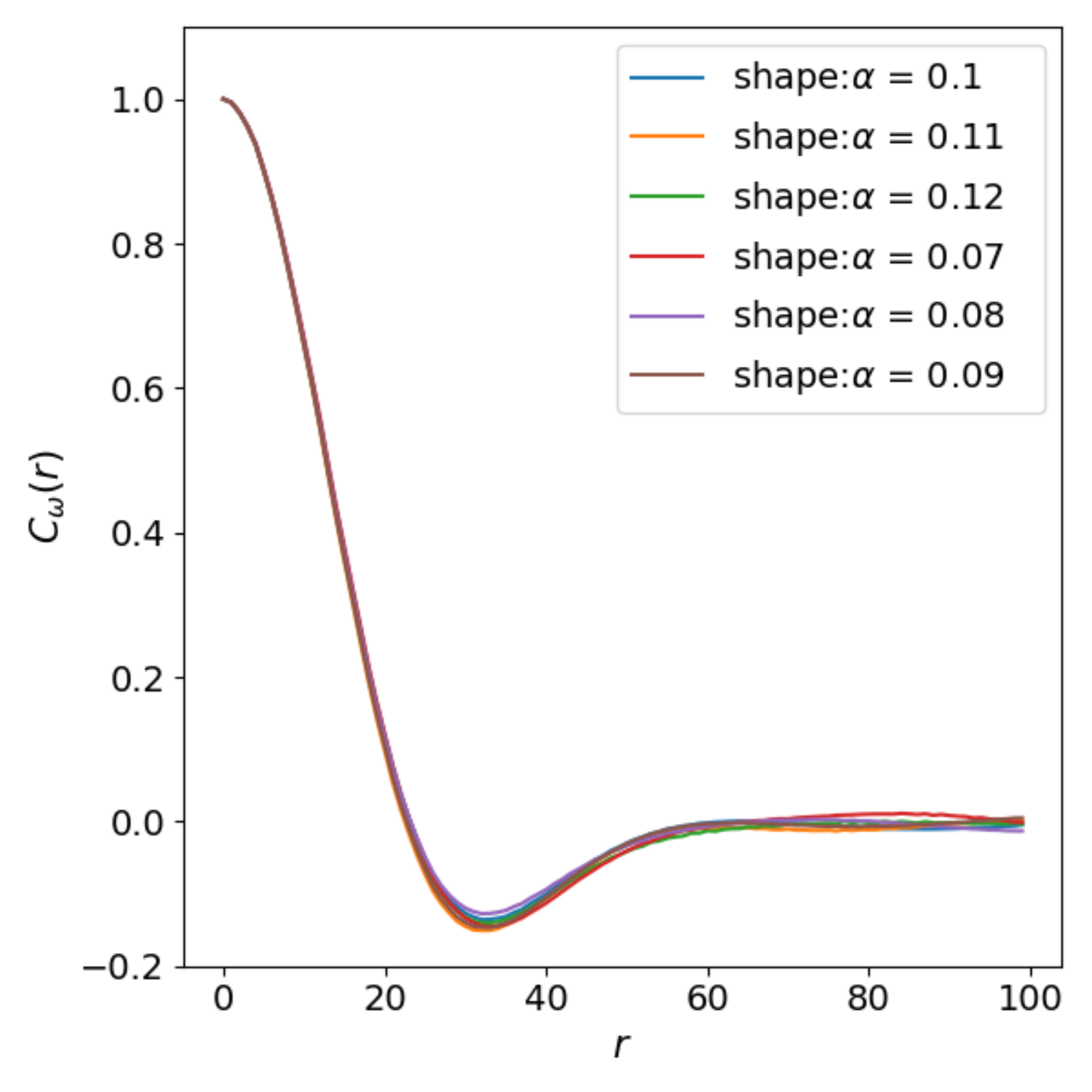}
			\caption{}
\includegraphics[width=155pt]{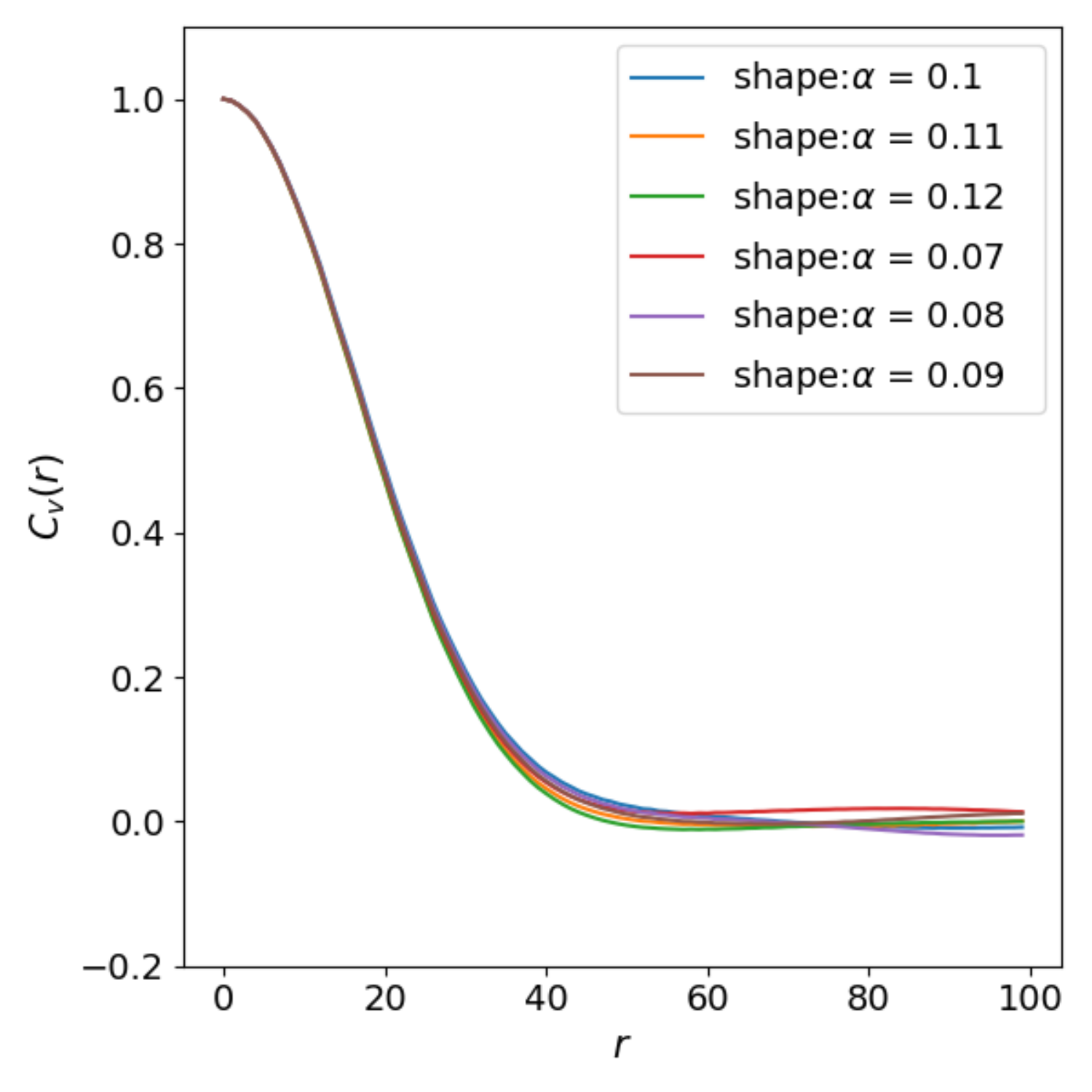}
	\caption{}
		\end{subfigure}
			\caption{{\bf Liquid state:} Snapshot showing a typical (a) velocity field, (b) vorticity field in the liquid state. (c) Vorticity-vorticity and (d) velocity-velocity correlation functions for $\alpha=0$ and different values of the intercellular driving $\zeta$. (e) Vorticity-vorticity and (f) velocity-velocity correlation functions  for $\zeta=0$ and different values of the polar driving $\alpha$. $\omega=0.005$.}
					\label{fig:3}
\end{figure}

We next investigate the properties of the liquid state in more detail. Fig.~\ref{fig:3}(a) shows a typical snapshot of the velocity field which is characterised by localised bursts of higher velocities, behaviour reminiscent of the active turbulence seen in active nematics~\cite{Doostmohammadi2018active}. Fig.~\ref{fig:3}(b) shows the vorticity field with vortices on the scale of a few cells. The other panels in the figure display this phenomenon more quantitatively by showing the vorticity-vorticity and velocity-velocity correlation functions for layers driven by inter-cellular driving (c), (d) or polar forcing (e),(f).   Correlations persist over $\sim 4$ cell diameters and are largely independent of the details or strength of the active forces suggesting that the range of the correlations is set by the passive forces. 
Movie 1, for $\alpha=0, \zeta=0.034$, and Movie 2, for $\alpha=0.062, \zeta=0$,  visually show the dynamics near the unjamming border for systems  driven solely by inter-cellular forces or polar forces respectively.  Polar forces result in larger velocities and there is a tendency of inter-cellular active forces to elongate the cells. Therefore we now measure the dependence of cell deformations on the active driving.

\subsection{Cell deformations}

In the absence of active forcing the cells will arrange into a honeycomb lattice and relax to a hexagonal shape. We investigate how they are stretched by the activity by measuring the average cell deformation,
\begin{equation}
D = \avg{\sqrt{{\mathcal D}_{xx,i}(t)^2 + {\mathcal D}_{xy,i}(t)^2}}{t,i} ,
 \end{equation}
where  ${\mathcal D}_{xx,i}$ and ${\mathcal D}_{xy,i}$ are the $xx$ and $xy$ components of the deformation tensor ${\mathcal D}$\ for cell $i$ and the average is taken over cells and time. Note that $D = 0.5$ corresponds to isotropic cells and $D=3$ to cells with an aspect ratio $\sim $1.75.

\begin{figure}[t!]
  	\begin{subfigure}[t]{0.24\textwidth}
  		\centering 
  		\includegraphics[width=115pt]{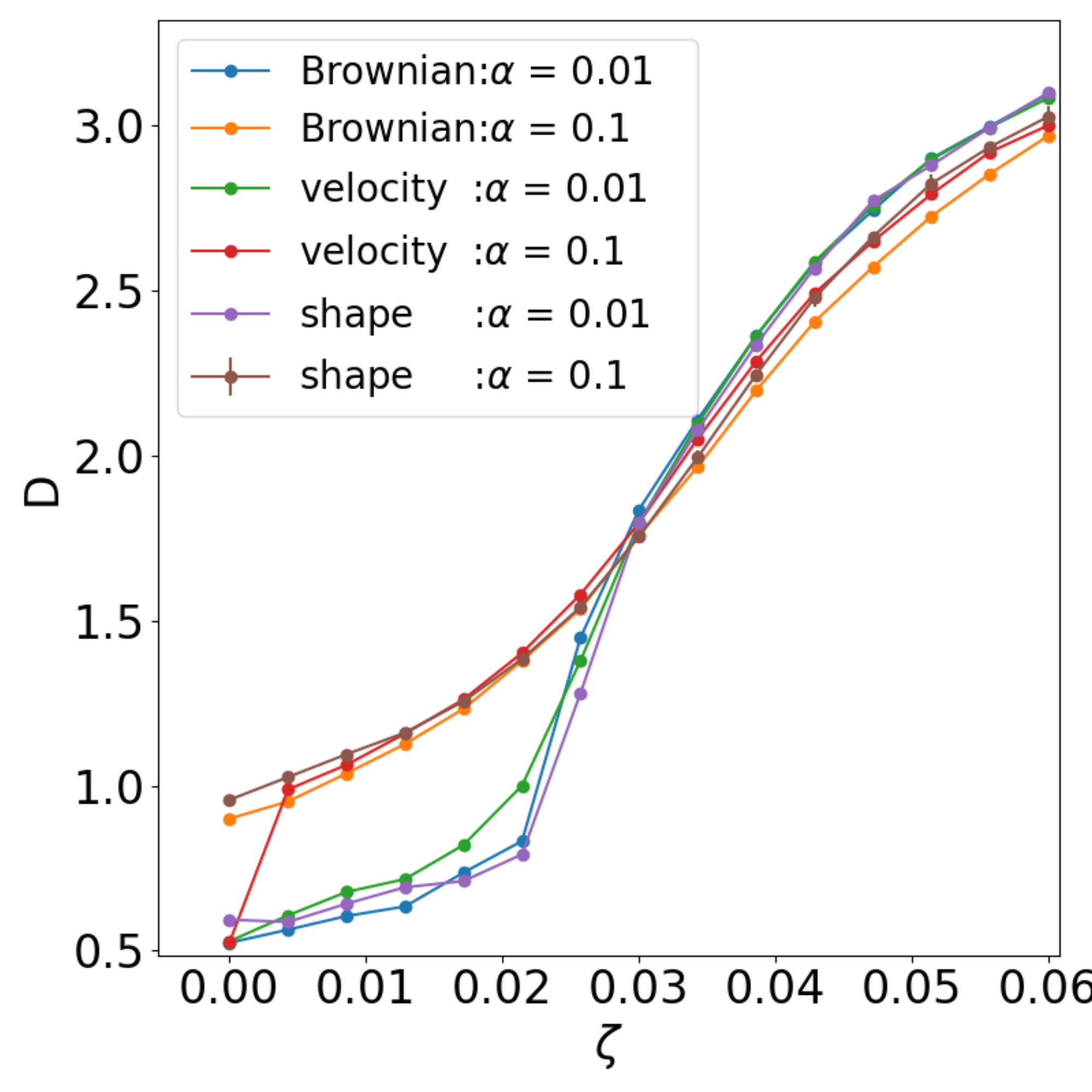}
  		\caption{}
		\label{fig:deformation_zeta}
  	\end{subfigure}
  	\begin{subfigure}[t]{0.24\textwidth}
  		\centering 
  		\includegraphics[width=115pt]{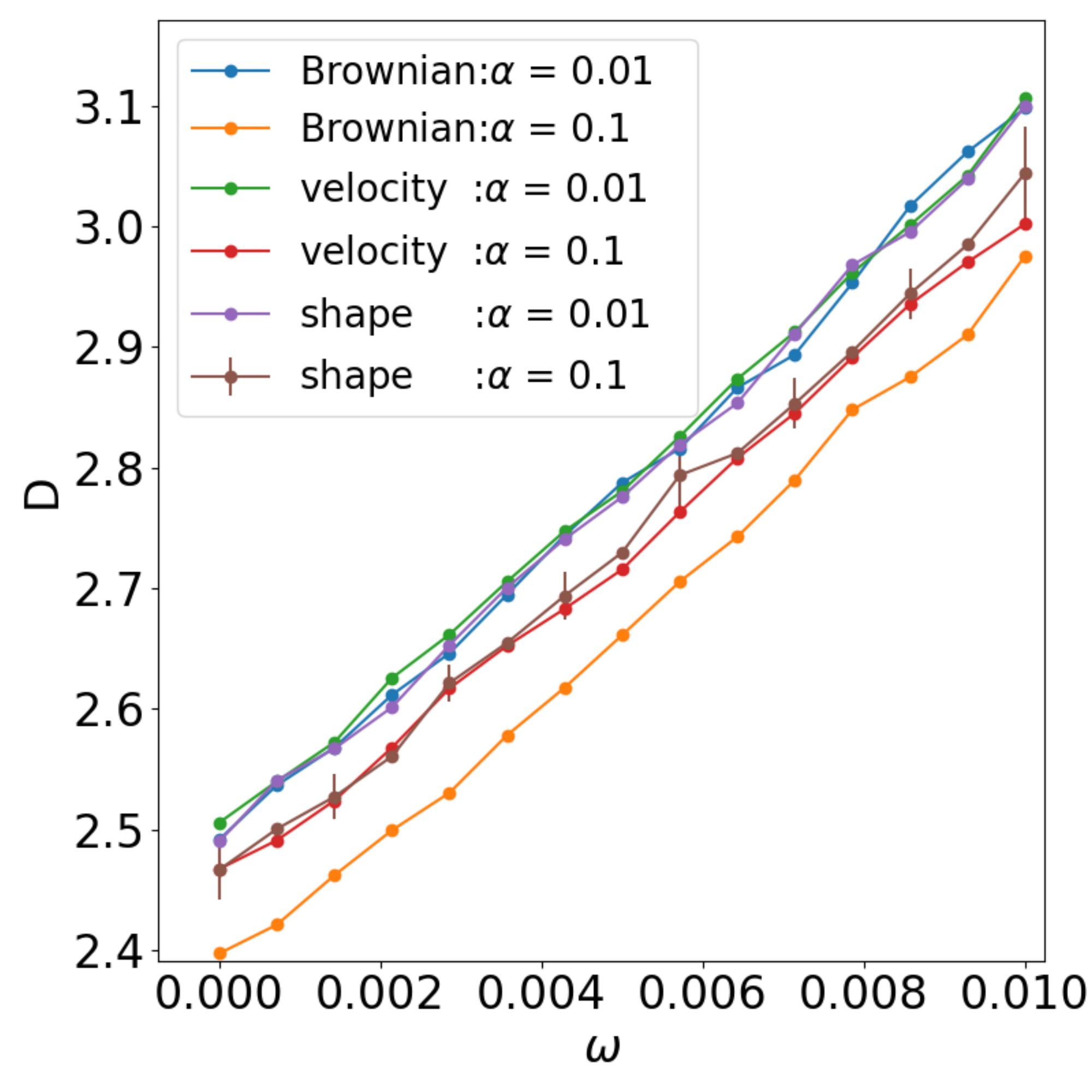}
  		\caption{ }
		\label{fig:deformation_adhesion}
  	\end{subfigure}
  	\begin{subfigure}[t]{0.24\textwidth}
  		\centering 
  		\includegraphics[width=115pt]{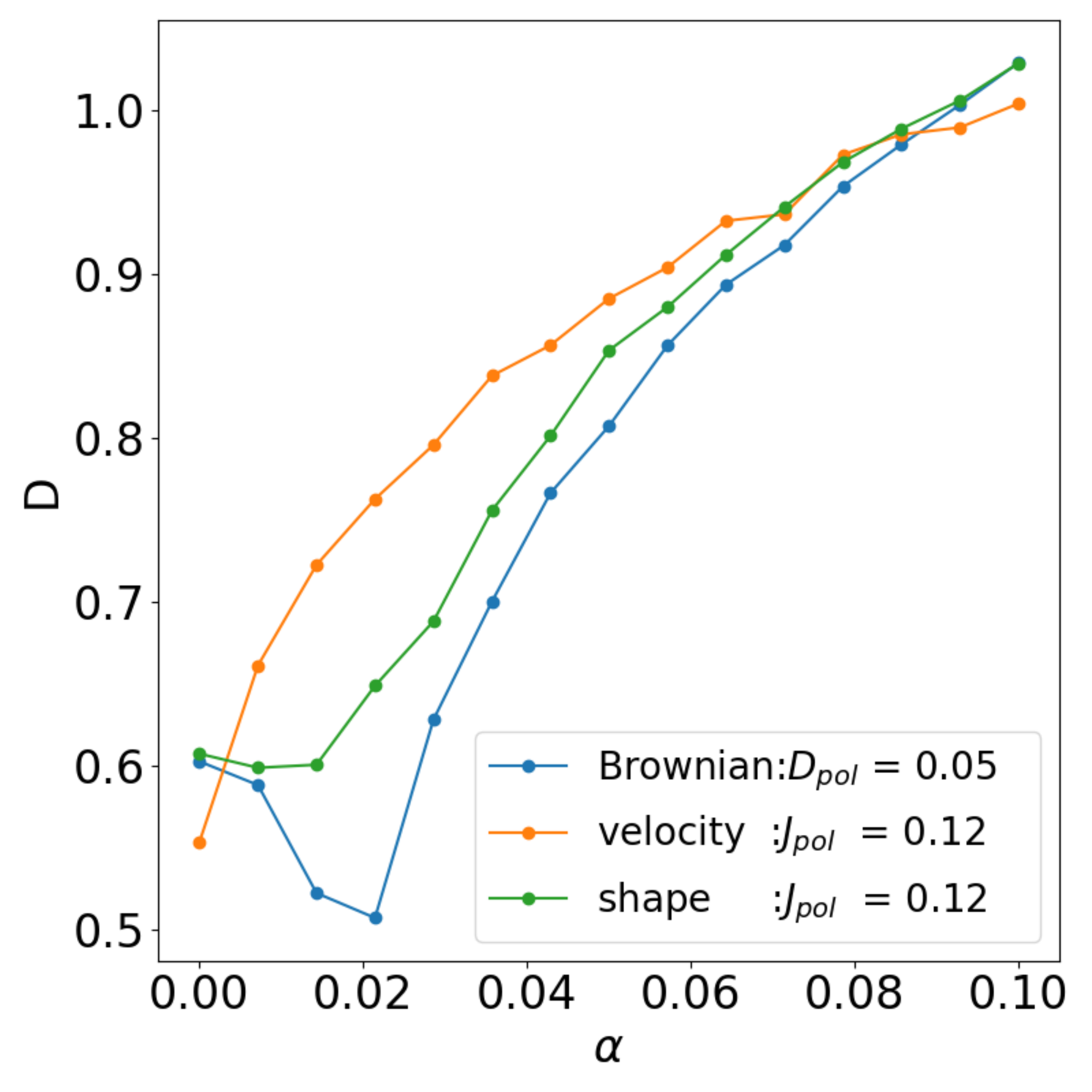}
  		\caption{}
		\label{fig:deformation_alpha1}
  	\end{subfigure}
  	\begin{subfigure}[t]{0.24\textwidth}
  		\centering 
  		\includegraphics[width=115pt]{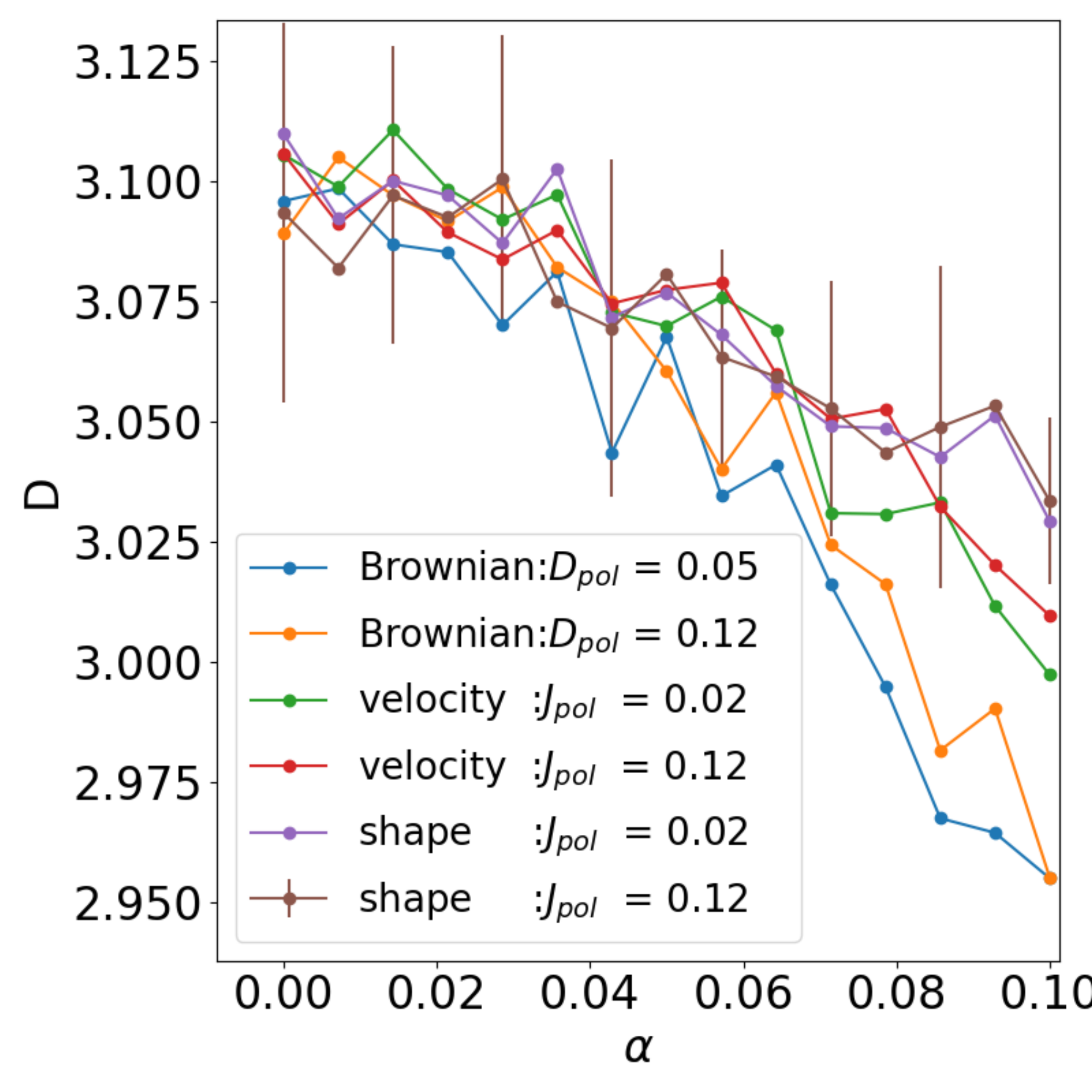}
  		\caption{}
			\label{fig:deformation_alpha2}
  	\end{subfigure}
  	\caption{{\bf Cell shape:} Average cell deformation $D$ as a function of (a) the  active inter-cellular force $\zeta$ for $\omega=0.01$; (b) the passive adhesion force  $\omega$ for $\zeta=0.06$, and the polar force $\alpha$ for (c) small active, inter-cellular forces $\zeta$=0.005, (d) large active, inter-cellular forces $\zeta=0.06$. $J_{\text{pol}}=0.12$ and $D_{\text{pol}}=0.12$ in (a) and (b). $\omega=0.01$ in (c) and (d).}
	\label{fig:deformation_alpha}
  \end{figure}

In all the models we consider changes in cell elongation are most marked in response to changes in the inter-cellular active forcing $\zeta$. For $\zeta>0$ (Fig.~\ref{fig:phasediagrama}), these  forces contribute to the deformation by elongating individual cells, and also by aligning neighbouring cells.
The variation of $D$ with $\zeta$ is displayed in Fig.~\ref{fig:deformation_zeta} showing that the dynamics of the polarisation is unimportant in  determining how the cells are stretched. Note that for small polar forces
 the increase of $D$ with $\zeta$ is particularly pronounced around $\zeta=0.025$ which corresponds to the crossover from the jammed to the liquid state.  Movies 1 and 3 compare the dynamics for $\alpha=0$, $\zeta=0.034$ and $\alpha=0$, $\zeta=0.044$ respectively showing visually the clear increase in cell extension with the active inter-cellular force.

The passive adhesion force between the cells also steadily increases the cell deformation (Fig.~\ref{fig:deformation_adhesion}) but to a lesser extent. This is because this term in the free energy (Eq.~\ref{adhesion}) acts as an effective negative line tension thus mimicking a preference of the cells to increase the length of cell-cell junctions.

Fig.~\ref{fig:deformation_alpha} also illustrates the changes in cell shape with the strength of the polar force $\alpha$. For small values of $\zeta$ where  deformation due to inter-cellular forces is small (Fig.~\ref{fig:deformation_alpha1}), the polar forces can themselves cause significant elongations of the cells. However, when $\zeta$ is large and the cells are already significantly stretched, adding polar forces has the opposite effect of reducing $D$ (Fig.~\ref{fig:deformation_alpha2}). This is because the cells tend to move apart more quickly, reducing their tendency to align with, and further extend, their neighbours.

Note that for small $\alpha$ the deformation can be greater than 0.5. This is because the cells remain stuck in the initial jammed state where some cells are deformed.  Weak active forces help the cells to move towards a state of hexagonal packing with minimum free energy and low deformation.

\subsection{Cell speed and flocking}

\begin{figure}[h!] 
  	\begin{subfigure}[t]{0.24\textwidth}
  		\centering 
			\includegraphics[width=115pt]{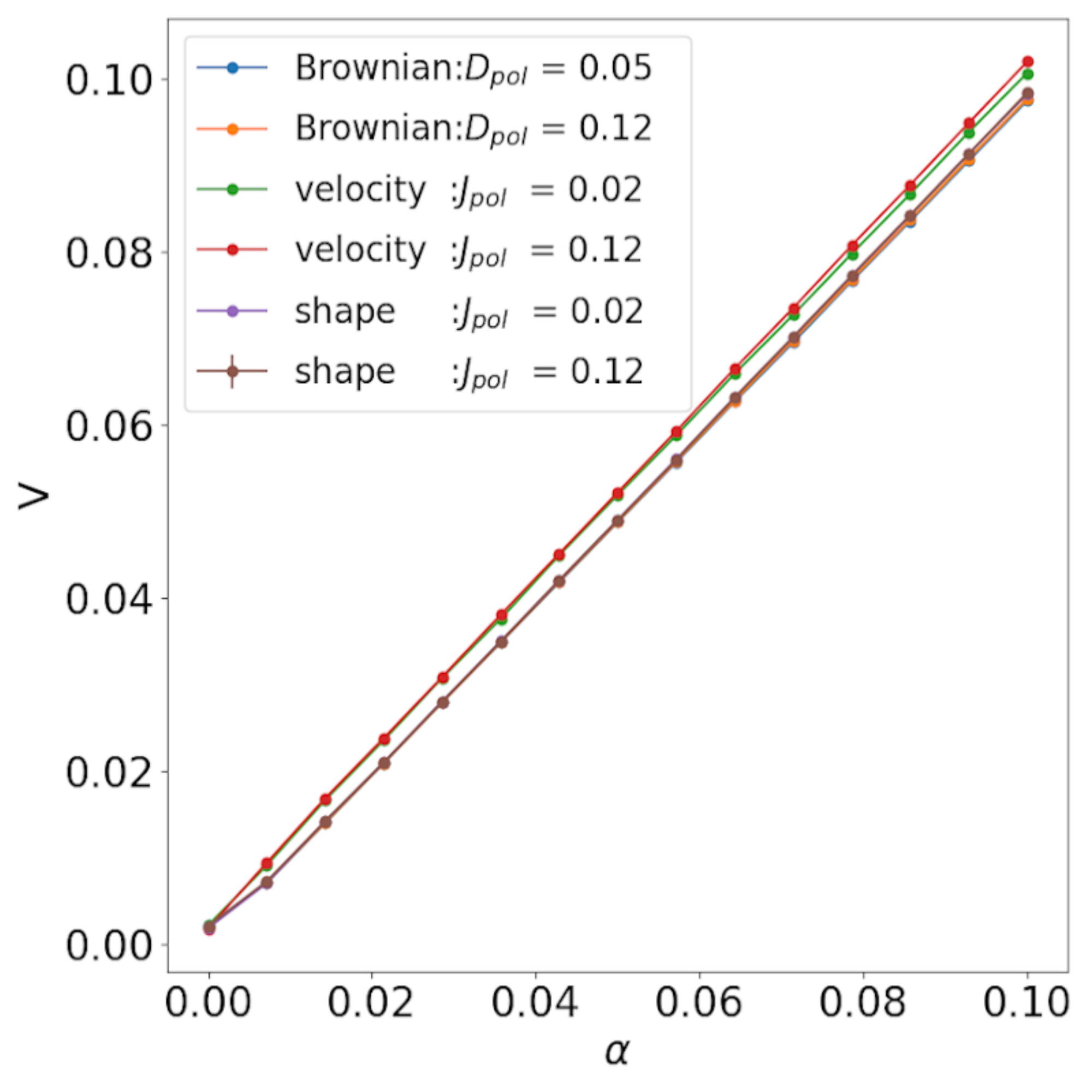}
  		\caption{}
		\label{fig:velocitya}
  	\end{subfigure}
  	\begin{subfigure}[t]{0.24\textwidth}
  		\centering 
		\includegraphics[width=115pt]{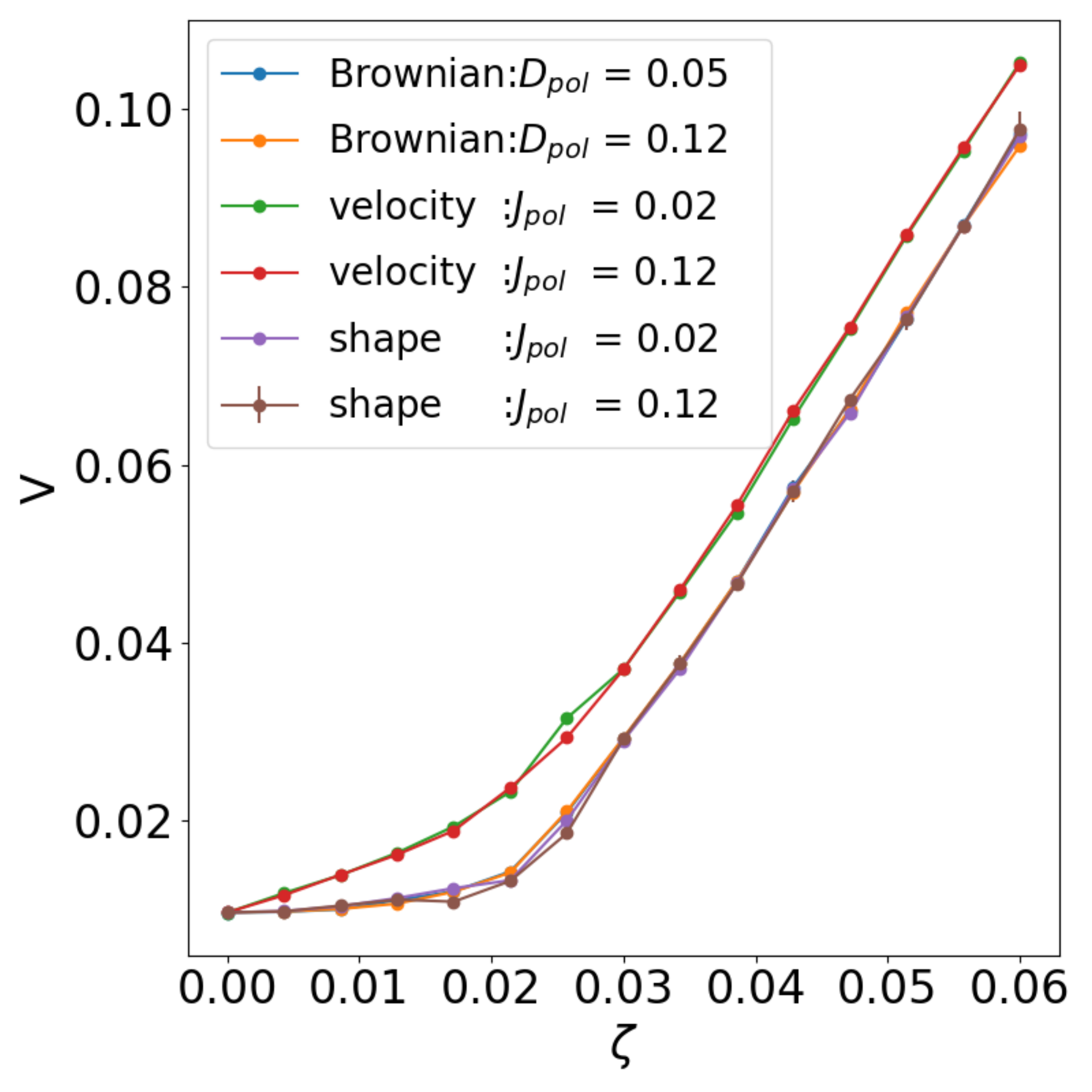}	
  		\caption{}
			\label{fig:velocityb}
  	\end{subfigure}
  	\begin{subfigure}[t]{0.24\textwidth}
  		\centering 
  	\includegraphics[width=115pt]{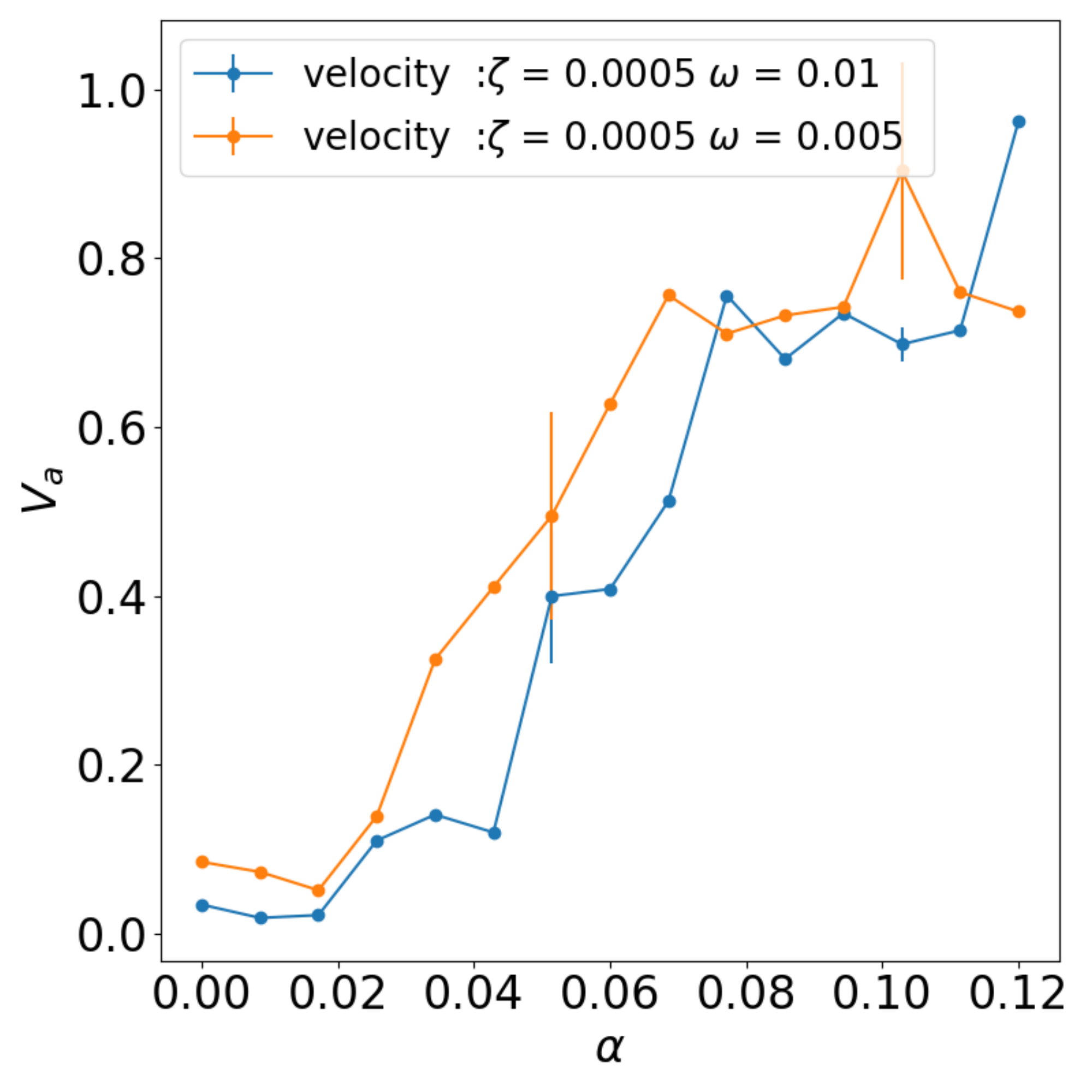}
  		\caption{}
  	\end{subfigure}
  	\begin{subfigure}[t]{0.24\textwidth}
  		\centering 
				\includegraphics[width=115pt]{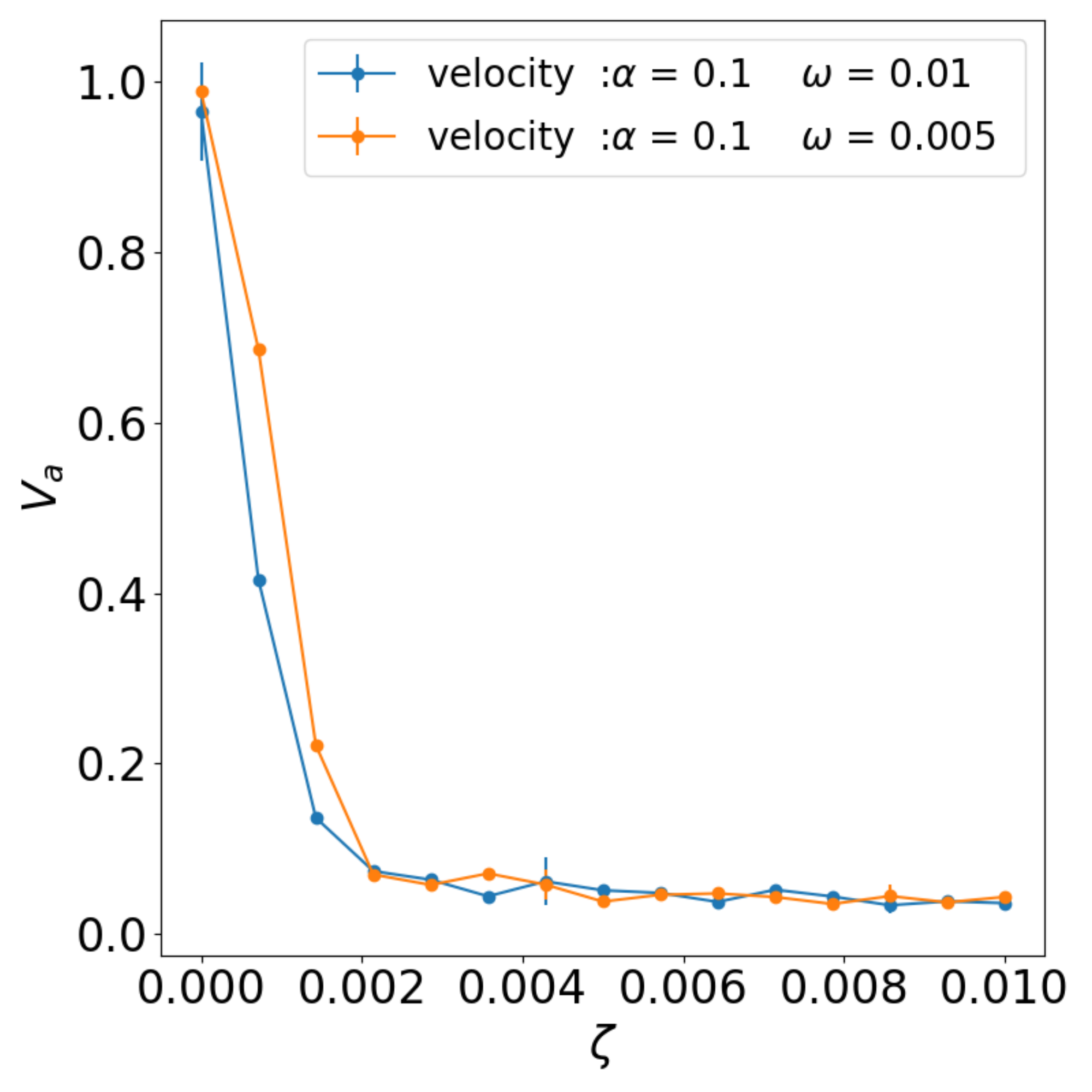}
  		\caption{}
  	\end{subfigure}
  	\caption{{\bf Cell movement:} Average cell speed $V$ as a function of the strength of (a) the polar force $\alpha$ for $\zeta=0.005$ and $\omega=0.01$, (b) the inter-cellular force $\zeta$ for $\alpha=0.01$ and $\omega=0.01$. Vicseck order parameter as a function of the strength of (c) the polar force $\alpha$, (d) the inter-cellular force $\zeta$. $J_{pol} = 0.12$ in both cases. The large error bars on the data in (c) are a consequence of the strong fluctuations in the flocking state.}
  	\label{fig:vicseck}
  \end{figure}

To further characterise the coordinated motion of cells  we next measured the averaged cell speed 
\begin{equation}
 V = \avg{\sqrt{v_{x,i}(t)^2 + v_{y,i}(t)^2}}{t,i}.
\end{equation}
Fig.~\ref{fig:velocitya} shows that the cell speed increases linearly with the strength of the polar force $\alpha$ as  expected. However, note that the cell speed also increases with the strength of the active inter-cellular forcing with a sharp change in the rate of increase around $\zeta \sim 0.25$, again corresponding to the transition from a jammed state to a liquid (Fig.~\ref{fig:velocityb}).

We recall the different equations of motion governing the direction of the cell polarisation, introduced in Sec.~\ref{sec:polalign}:  {\it shape}, relaxing to the long axis of the cell; {\it velocity}, relaxing to the direction of the cell velocity; and  {\it Brownian}, undergoing a persistent random walk. The cell speed, and indeed all other parameters we have considered so far, have had only a small quantitative dependence on the definition used. However, as the strength of the polar force is increased, a new dynamical phase appears, just for the case {\it velocity}. This is  flocking, where large numbers of neighbouring cells move in the same direction. 

To quantitatively characterise the flocking behaviour we define the Vicseck order parameter
\begin{equation}
V_a = \left\langle \frac{1}{N} \left| \sum_{i}  \frac{\mathbf{v}_i^{\text{COM}}}{\left| \mathbf{v}_i^{\text{COM}} \right|} \right| \right \rangle_{t}
\end{equation}
where $V_a = 1$ corresponds to all the cells moving in the same direction and $V_a = 0$ for random motion. Fig.~\ref{fig:vicseck}c shows that flocking occurs as the polar force increases. However, for a given $\alpha$, Fig.~\ref{fig:vicseck}d indicates that inter-cellular driving acts to destroy the flocking. 
This occurs because, as the inter-cellular active forces increase, the overall repulsive interaction between cells which is needed to stabilise flocks \cite{deseigne2010collective,grossman2008emergence} becomes weaker because the effective adhesion arising from the inter-cellular interactions becomes stronger. As a result, the flocking state disappears, but there are still local clusters inside which the velocities of cells align.

Fig.~\ref{fig:phasediagramd} shows the phase diagram for the case  {\it velocity} where the polarisation aligns to the direction of the cell velocity. We define flocking states as those with a Viscek order parameter $V_a > 0.15$. The flocking state can be either liquid-like where cells rearrange within the flock or, for larger polar forces, solid-like where they retain their positions.
Movie 4 is an example of flocking for $\alpha=0.08$ and $\zeta=0$.

\section{Discussion}

We have used a phase-field approach to model the collective motion of a confluent layer of cells. The phase-field model includes passive forces, which minimise a free energy, as well as active forces, which drive the system out of equilibrium.
Using our model, we investigate the relative importance of active polar forces, resulting from lamellopodia pulling on the substrate, with active inter-cellular forces, resulting from force transmission across cell junctions, in determining the cell dynamics.
We compare different possibilities for defining the dynamics of the cell polarisation: {\it shape}, relaxing to the long axis of the cell; {\it velocity}, aligning to the direction of the cell velocity; and  {\it Brownian}, undergoing a persistent random walk.

Independent of the way in which the polarity is implemented, a predominant feature of the dynamics is a crossover from a jammed state, where any cell rearrangements are strongly suppressed, to a liquid, where the cells can move relative to each other \cite{bi2016motility}.  The transition can be driven either by polar or by inter-cellular forces, and in both cases the liquid shows localised bursts of velocity as well as vorticity correlations over several cells, which are phenomena characteristic of the active turbulence observed in active nematics \cite{Doostmohammadi2018active}. Correlations are largely independent of the details or strength of the active forces, suggesting that they are  mainly controlled by passive forces and are rather insensitive to the exact form of the active driving.

Surprisingly, different mechanisms of polarity alignment not only result in similar patterns of collective motion, but even lead to close quantitative measures of cell rearrangement rates, averaged cell deformation, and averaged cell velocity in the monolayer. However, a striking exception is when the polarity aligns to the cell velocity, in which case the cells transition to a flocking state at strong polar forcing. The flocking state is destroyed as the strength of the inter-cellular forces is increased. This agrees well with the experimental results presented in~\cite{kim2013propulsion}, suggesting that for human mammary epithelial MCF-10A cells the polarity of the cells aligns closely to their velocities.

The way we choose to implement the forcing in the model, and the parameter values we consider, will undoubtedly need modification in the light of future experimental results. For example, we have chosen a polar force that is uniformly distributed over a cell, and it would be interesting to investigate how moving the centre of mass of this force distribution towards the front of a cell changes the dynamics. We have concentrated on inter-cellular forcing that is extensile, i.e.~that acts to increase the extension of a cell and will, in future, consider the contractile case where the active inter-cellular forces act in concert with the surface tension to decrease cellular distortion. We would also like to further investigate the role of inter-cellular adhesion in the dynamical behaviour of the layer, and move away from the condition of confluency.

The force distribution within cell monolayers has been measured experimentally by several authors using traction force microscopy and monolayer stress microscopy~\cite{tambe2011collective,kim2013propulsion,tambe2013monolayer}. However, it is not yet clear how the forces should be divided into passive or active, or into polar and inter-cellular contributions. Indeed, it would be of particular interest to vary the polar forces acting on cells by varying the stiffness of substrates \cite{sunyer2016collective} or using drug treatments on the kinases such as ROCK which are involved in actin filament protrusion \cite{ladoux2017mechanobiology}.  The strength of inter-cellular  stresses could be modified by controlling the levels of mechanosensors such as $\alpha$-catenin at the cell-cell junctions \cite{saw2017topological}. 
Comparing the force density in different situations, both experimentally and in models such as the one presented here, will provide a framework which will help to interpret current and future measurements, leading to an improved understanding of tissue mechanics in confluent cell layers and epithelia.

 \section*{Supplementary Movies}
 \begin{itemize}
	\item{Movie 1: Phase field simulation of cell dynamics in the liquid phase near the unjamming boundary driven by inter-cellular forces ($\alpha = 0.0,\zeta=0.034$). The polarization relaxes towards the long axis of the cell (case {\it shape}). There are 800 cells in a $200\times200$ domain and red arrows indicate the centre of mass velocity for each cell.}
	
	\item{Movie 2: Phase field simulation of cell dynamics in the liquid phase near the unjamming boundary driven by polar forces ($\alpha = 0.062,\zeta=0.0$). The polarization relaxes  
towards the long axis of the cell (case {\it shape}). There are 800 cells in a $200\times200$ domain and red arrows indicate the centre of mass velocity for each cell.}
	
	\item{Movie 3: Phase field simulation of cell dynamics in the liquid phase driven by inter-cellular forces ($\alpha = 0.0,\zeta=0.044$). The polarization relaxes towards the long axis of the cell (case {\it shape}). There are 800 cells in a $200\times200$ domain and red arrows indicate the centre of mass velocity for each cell.}

	\item{Movie 4: Phase field simulation of cell dynamics showing flocking ($\alpha = 0.08,\zeta=0.0$). The polarization relaxes towards the velocity of the cell (case {\it velocity}). There are 233 cells in a $108\times108$ domain and red arrows indicate the centre of mass velocity for each cell.}
\end{itemize}

\section*{Acknowledgements}
A. D. acknowledges support from the Novo Nordisk Foundation (grant No.~NNF18SA0035142) and funding from the European Union's Horizon 2020 research and innovation program under the Marie Sklodowska-Curie grant agreement No.~847523 (INTERACTIONS). G. Z. acknowledges funding support from the China Scholarship Council(grant agreement No. 201808220082) and the University of Oxford.\\

\end{document}